\documentclass[twocolumn,showpacs,preprintnumbers,amsmath,amssymb]{revtex4-1}
\usepackage{epsfig}
\usepackage{graphicx}
\usepackage{epsfig}
\usepackage{rotating}
\usepackage{amssymb}
\usepackage{amsmath}
\usepackage{graphicx}
\usepackage{amssymb}
\usepackage{amsmath}
\usepackage{color}

\def\beq{\begin{eqnarray}}
\def\eeq{\end{eqnarray}}

\newcommand{\be}{\begin{equation}}
\newcommand{\ee}{\end{equation}}
\newcommand{\bea}{\begin{eqnarray}}
\newcommand{\eea}{\end{eqnarray}}

\begin{document}

\title{Advective-diffusive motion on large scales from small-scale dynamics with an internal symmetry}

\author{Raffaele Marino}
 \email{rmarino@kth.se}
\affiliation{NORDITA, Royal  Institute  of  Technology  and  Stockholm  University, Roslagstullsbacken  23,  SE-106  91  Stockholm,  Sweden}
\author{Erik Aurell}
\affiliation{Dept. of Computational Biology and Center for Quantum Materials, KTH -- Royal Institute of Technology,  AlbaNova University Center, SE-106 91~Stockholm, Sweden}
\altaffiliation{Depts. Information and Computer Science and Applied Physics, Aalto University, Espoo, Finland}

\begin{abstract}
We consider coupled diffusions in $n$-dimensional space and on a compact manifold
and the resulting effective advective-diffusive motion on large scales in space. The effective drift
(advection) and effective diffusion are determined as a solvability conditions in a multi-scale analysis.
As an example we consider coupled diffusions in $3$-dimensional space and on the group manifold $SO(3)$
of proper rotations, generalizing results obtained by H. Brenner (1981). We show in detail how
the analysis can be conveniently carried out using local charts and invariance arguments.
As a further example we consider coupled diffusions in $2$-dimensional complex space and on the group manifold $SU(2)$.
We show that although the local operators may be the same as for $SO(3)$, due to the global
nature of the solvability conditions the resulting diffusion will be different, and generally more isotropic.  
\end{abstract}

%\pacs{?????}

%\keywords{????}
\maketitle

%%%%% -------------------------------------------------------------------------
%%%%% -------------------------------------------------------------------------
%%%%% -------------------------------------------------------------------------

\section{Introduction}
\label{sec:introduction}
The transition from the microscopic to the macroscopic is 
the central problem in statistical physics~\cite{Feynman}.
Posed already in the 19th century it can be understood as the emergence of qualitatively different phenomena 
on larger scales from averaging of (typically simpler) phenomena on smaller scales, cf~\cite{ChibbaroRondoniVulpiani,On3Levels,GrootMazur-book}.
Analogous considerations were later extended to hydrodynamics~\cite{Speigel81} 
and more general systems of ordinary or partial differential equations
and are today usually referred to as homogenization~\cite{BLP78}, or as multi-scale methods~\cite{PavliotisStuart08}.

In this paper we consider effective diffusion on large scales from averaging motion involving additional
degrees of freedom on small scales. Our setting is mesoscopic physics where the small scale motion
is understood as an overdamped dynamics with given transport coefficients, and where the large-scale motion is 
described by other, effective, transport coefficients. 
As will be described in detail below we thus apply the multi-scale method to the analysis
of overdamped Fokker-Planck equations (FPE) where we start with an FPE on configuration space
and an internal symmetry, and end up with an FPE on configuration space only. The
effective transport coefficients on large scales are computed as averages over the small scales.

In a recent joint work with Bo, Dias and Eichhorn,
we considered the instance of the problem where the configuration space is ordinary three-dimensional
space and the internal symmetry the group of rotations $SO(3)$~\cite{5Authors}. We were thus able
to recover results obtained by H.~Brenner 35 years ago~\cite{Brenner1981}. The most striking of these results
is a contribution to the effective diffusivity which is inversely proportional to $k_BT$ and which
depends quadratically on the applied force field. As stressed already by Brenner this effect should completely
dominate the diffusion of \textit{e.g.} non-spherical micron-sized particles settling in a gravitational field,
and is thus quite important for the understanding of sedimentation and aggregation during sedimentation.
Further physical consequences of this theory will be discussed in a forthcoming separate contribution~\cite{BoEichhornMarino}.

The objectives of the present paper are threefold. 
First we will separate the different strands of the analysis with the aim to more clearly distinguish the technical
and conceptual points. We will thus show that advection-diffusion on large scales 
is quite general and holds if the internal symmetry is described by a compact manifold of otherwise
arbitrary internal geometry. The steps in the multi-scale analysis can be posed as solving elliptic partial 
differential equations on the manifold, and then computing weighted averages of these solutions.   
These are both well-defined, though generally non-trivial tasks, consequently it is only
determining the effective transport coefficients in closed form which
necessitates additional assumptions.
Second we will give further details on the analysis  
in~\cite{5Authors} focusing on a local representation of rotations (elements of $SO(3)$) as charts,
and on the computation of the effective transport coefficients 
by the use of $SO(3)$ invariant theory.
We will also present results going beyond a diagonalizability 
assumption made in~\cite{5Authors}, predictions
which we compare to numerical simulations.
Third we will show that the first step of the multi-scale analysis can 
be carried out also for a process that satisfies a detailed balance condition. In this case, however, the transport coefficients
cannot be analytically computed as averages of known terms over the manifold using simply invariant theory.
%except for the case of consecutive
%uniform measure these average however cannot be simply
%computed by invariance theory.

As an illustration we also include a discussion of the example where the configuration space is $\mathbb{C}^2$, two-dimensional 
complex space, and the internal symmetry is the Lie group $SU(2)$ of $2\times2$ unitary matrices of unit determinant.
The small-scale motion is thus coupled advection-diffusion on $\mathbb{C}^2$ and on the group manifold of $SU(2)$ while the large scale motion is
advection-diffusion on  $\mathbb{C}^2$ only. As a consequence of the Lie algebras $su(2)$ and $so(3)$ being isomorphic
the equations to solve on the consecutive levels of the multi-scale analysis are equivalent to those for $SO(3)$,
but the solvability conditions, being global, are different. 

The paper is organized as follows. In Section~\ref{s:diffusion-internal}
we introduce our basic model of coupled diffusions in space and on a manifold
and the basic notions of differential geometry which we need in the following.
In Section~\ref{s:multi-scale-internal} we carry out a multi-scale analysis
on the basic model and shows that it leads to large-scale drift and diffusion.
In Section~\ref{s:internal-symmetry} we apply the basic model to motion in space and a symmetry group
acting on that space, and in Section~\ref{s:so3} we discuss the
example where space is three-dimensional and the symmetry is rotation group $SO(3)$.
In Section~\ref{s:su2} we similarly discuss the case of  $SU(2)$, in Section~\ref{s:secsim}
we present numerical results, and in Section~\ref{s:discussion} we sum up and discuss
what has been done. Some standard material is for completeness included as appendices.

\section{Diffusion with internal states}
\label{s:diffusion-internal}
In this section we introduce our basic model, coupled diffusion on 
$n$-dimensional space and on a manifold $\mathcal{M}$ with dimension $m$.
The general theory of Brownian motion on manifolds was developed by Kolmogorov, It\^o and Yosida and others 
in the middle of the 20th century, and
%was constructed more than 60 years ago by
%It\^o\cite{Ito1950} and Yosida~\cite{Yosida1952}, and is described in recent monographs \textit{cf}~\cite{Hsu2002}
is described in many monographs \textit{cf}~\cite{McKean67book,BookSPLG1,BookSPLG2,Hsu2002}.
The case of Brownian motion on $SO(3)$, which will be one of our main examples, was
explicitly constructed by McKean in an early but still very instructive paper~\cite{McKean1960}.

\subsection{Brownian motion on manifolds}
Our purpose in this Section is to introduce the notation and set the stage for
the general multi-scale analysis in the following Section~\ref{s:multi-scale-internal}.
We therefore start from the dictionary definition that
an $m$-dimensional manifold is a topological space, each point of which has a neighbourhood that is homeomorphic to the Euclidean space of dimension $m$. 
This means that the manifold can be covered by a collection of open sets $U_i$ which are one-to-one and smoothly related to open sets $V_i\in \mathbb{R}^m$.
One set $V_i$ is called a local coordinate patch for the set $U_i$, and the map $\psi_i:U_i\to V_i$ is called a local coordinate. 
Suppose a point $p$ on the manifold belongs to two sets $U_i$ and $U_j$ and $U'$ is an open set in $U_i \cap U_j$ containing $p$.
Then we can define two sets $V'_i=\psi_i(U')$ and $V'_j=\psi_j(U')$ and two maps $\psi_{ij}=\psi_j\circ\psi_i^{-1}:V'_i\to V'_j$ and  $\psi_{ji}=\psi_i\circ\psi_{j}^{-1}:V'_j\to V'_i$
which have the meaning of a change of coordinate, locally around point $p$. Obviously $\psi_{ji}=\psi_{ij}^{-1}$.

On the manifold is defined a metric $g$ which can be expressed as a matrix $g_{ij}$ in local coordinates. This means that if 
two points $p$ and $q$ are close and have coordinates $\alpha_p=\psi_i(p)$ and $\alpha_q=\psi_i(q)$ in the same patch such that $\alpha_q=\alpha_p+\Delta \alpha$
then the squared distance $d^2(p,q)$ equals $\sum_{ab} g_{ab}(\alpha) \Delta \alpha^a  \Delta \alpha^b$. Under a change of coordinate $\alpha\to \alpha'$
with Jacobian $J^a_b=\frac{\partial \alpha'^a}{\partial \alpha^b}$ the metric
$g$ therefore transforms as a second-order contravariant tensor
\begin{equation}
g'_{ab}(\alpha')=\sum_{cd} (J^{-1})^c_a  (J^{-1})^d_b  g_{cd}(\psi_{ji}(\alpha')), \nonumber
\end{equation}
The manifold also has a volume element $\sqrt{\det g}$, which we will write $\sqrt{g}$, and  which under a coordinate change transforms as $\sqrt{g'}=\sqrt{g}/|\det J|$.
We will from now on use the Einstein convention where repeated indices, one upper and one lower, are summed. 
 
Let now $\mathbb{R}^n\otimes V_i$ be parametrized as $(x^1,\ldots,x^n,\alpha^1,\ldots,\alpha^m)$. The building blocks of our basic model are systems of coupled stochastic differential
equations
\begin{widetext}
\begin{equation}\label{Overdampedequations}
\begin{split}
& \text{d}x^{i}= A_1(\vec{x},\vec{\alpha},t)^{i}\text{d}t+B_{11}(\vec{x},\vec{\alpha},t)^{i}_j\bullet\text{d}W^j+B_{12}(\vec{x},\vec{\alpha},t)^{i}_b\bullet\text{d}{W'}^b,\\
& \text{d}\alpha^{a}=A_2(\vec{x},\vec{\alpha},t)^{a} \text{d}t+B_{21}(\vec{x},\vec{\alpha},t)^{a}_j\bullet\text{d}W^j+B_{22}(\vec{x},\vec{\alpha},t)^{a}_b\bullet\text{d}{W'}^{b},
\end{split}
\end{equation}
\end{widetext}
where $\text{d}W^j$ and $\text{d}{W'}^b$ are independent standard $n$-dimensional and $m$-dimensional Wiener noises, $\vec{A}_1$ and $\vec{A}_2$ 
$n$- and $m$-dimensional vector fields and $B_{11}$, $B_{12}$, $B_{21}$ and $B_{22}$ are respectively $n\times n$, $n\times m$, $m\times n$ and $m\times m$-dimensional matrix fields.
All functions are assumed to depend smoothly on their arguments. The It\^o prescription is used.
The generator of the diffusion is the operator
\begin{widetext}
\begin{equation}\label{generator of diffusions}
\mathcal{L}=A_1(\vec{x},\vec{\alpha},t)^{i} \frac{\partial}{\partial x^i} +A_2(\vec{x},\vec{\alpha},t)^{a}\frac{\partial}{\partial  \alpha^a}+\frac{1}{2}b_{11}(\vec{x},\vec{\alpha},t)^{ij}\frac{\partial^{2}}{\partial x^i \partial x^j}+\frac{1}{2}b_{22}(\vec{x},\vec{\alpha},t)^{ab} \frac{\partial^{2}}{\partial \alpha^{a} \partial \alpha^{b}}+b_{12}(\vec{x},\vec{\alpha},t)^{ia} \frac{\partial^{2}}{\partial x^{i} \partial \alpha^{a}},
\end{equation}
\end{widetext}
with
\begin{eqnarray}
b_{11}^{ij}=\left(B_{11} B_{11}^T + B_{12} B_{12}^T\right)^{ij} \nonumber ;\\ 
b_{22}^{ab}=\left(B_{22} B_{22}^T + B_{21} B_{21}^T\right)^{ab} \nonumber; \\
b_{12}^{ia}= \left(B_{11} B_{21}^T  + B_{12} B_{22}^T\right)^{ia} \nonumber.
\end{eqnarray}
The generator determines the time change of expectation values of future events through $\partial_t E + \mathcal{L}E=0$.
The evolution of a probability density is given by the Fokker-Planck equation $\partial_t P = \mathcal{F}P$, where
\begin{widetext}
\begin{eqnarray}\label{Fokker-Planck-operator}
\mathcal{F}P &=&  -\frac{\partial}{\partial x^i}\left(A_1(\vec{x},\vec{\alpha},t)^{i}P\right) - \frac{\partial}{\partial  \alpha^a}\left(A_2(\vec{x},\vec{\alpha},t)^{a}P\right) 
\nonumber \\
&& +\frac{1}{2}\frac{\partial^{2}}{\partial x^i \partial x^j}\left(b_{11}(\vec{x},\vec{\alpha},t)^{ij}P\right)
+\frac{1}{2}\frac{\partial^{2}}{\partial \alpha^{a} \partial \alpha^{b}}\left(b_{22}(\vec{x},\vec{\alpha},t)^{ab}P\right) 
+\frac{\partial^{2}}{\partial x^{i} \partial \alpha^{a}}\left(b_{12}(\vec{x},\vec{\alpha},t)^{ia} P\right).
\end{eqnarray}
\end{widetext} 
Note that being a density, $P$ transforms as $P'=P/|\det J|$.

During a sufficiently short time interval the solutions of (\ref{Overdampedequations}) will with high probability not leave $\mathbb{R}^n\otimes V_i$ and then specify a trajectory 
in  $\mathbb{R}^n\otimes \mathcal{M}$. Assume that the trajectory is contained in a set $\mathbb{R}^n\otimes U'$ centered around point $p$ 
parametrized by two coordinate systems $(\alpha^1,\ldots,\alpha^m)\in V_i$ and  $(\alpha'^1,\ldots,\alpha'^m)\in V_j$ with Jacobian $J$. 
The  It\^o lemma specifies the motion in the second coordinate system to be
\begin{widetext}
\begin{equation}\label{Overdampedequations-primed}
\begin{split}
& \text{d}x^{i}= {A'}_1(\vec{x},\vec{\alpha'},t)^{i}\text{d}t+{B'}_{11}(\vec{x},\vec{\alpha'},t)^{i}_j\bullet\text{d}W^j+{B'}_{12}(\vec{x},\vec{\alpha'},t)^{i}_b\bullet\text{d}{W'}^b,\\
& \text{d}{\alpha'}^{a}={A'}_2(\vec{x},\vec{\alpha'},t)^{a} \text{d}t+{B'}_{12}(\vec{x},\vec{\alpha'},t)^{a}_j\bullet\text{d}W^j+{B'}_{22}(\vec{x},\vec{\alpha'},t)^{a}_b\bullet\text{d}{W'}^{b},
\end{split}
\end{equation}
\end{widetext}
where
\begin{equation}
\label{eq:transformations}
\begin{split}
& {A'}_1^i =   A_1(\vec{x},\alpha(\vec{\alpha'}),t)^i;\\
& {A'}_2^a =   J^a_b A_2(\vec{x},\alpha(\vec{\alpha'}),t)^b + \frac{1}{2}\left(\frac{\partial^2 {\alpha'}^a}{\partial \alpha^c \partial \alpha^d}\right) b_{22}^{cd};\\
& {{B'}_{11}}^{i}_j =   B_{11}(\vec{x},\alpha(\vec{\alpha'}),t)^{i}_j;\\
& {{B'}_{12}}^{i}_b =   B_{12}(\vec{x},\alpha(\vec{\alpha'}),t)^{i}_b;\\
& {{B'}_{21}}^{a}_i =   J^a_c B_{21}(\vec{x},\alpha(\vec{\alpha'}),t)^{c}_i;\\
& {{B'}_{22}}^{a}_b =   J^a_c B_{22}(\vec{x},\alpha(\vec{\alpha'}),t)^c_b.\\
\end{split}
\end{equation}
from what follows
\begin{equation}
\label{eq:more-transformations}
\begin{split}
& {{b'}_{11}}^{ij} =   b_{11}^{ij};\\
& {{b'}_{12}}^{ia} =   J^{a}_c b_{12}^{ic};\\
& {{b'}_{22}}^{ab} =   J^a_c J^b_d b_{22}^{cd}.\\
\end{split}
\end{equation}
It is readily checked that the solutions of the Fokker-Planck equation (\ref{Fokker-Planck-operator}) change by the transformations (\ref{eq:transformations}) and (\ref{eq:more-transformations}).

Diffusion on $\mathbb{R}^n\otimes \mathcal{M}$ is specified as a family of systems of stochastic differential
equations (\ref{Overdampedequations}) in local coordinates which transform as 
(\ref{Overdampedequations-primed}) and
(\ref{eq:transformations}) and (\ref{eq:more-transformations})
under any well-defined change of variables whatsoever.

\subsection{Integration, Adjoint and Fokker-Planck}

Integration on $\mathcal{M}$ is defined by first dividing it up in sets $B_i$ each contained in an open set $U_i$ with local coordinates $(\alpha_i^1,\ldots,\alpha_i^m)$
in local patch $V_i$. Let the metric $g$ expressed in the coordinates of local patch $V_i$ be the matrix-valued function $g_{ab}^{(i)}(\alpha_i)$ and let $f$
be a function $f:\mathcal{M}\to \mathbb{R}$. Then 
\begin{equation}
\int \sqrt{g} f = \sum_i \int_{\psi_i(B_i)}\sqrt{g^{(i)}} d\alpha_i^1\cdots d\alpha_i^m f(\psi_i^{-1}(\alpha_i)).
\end{equation}
Integrating two functions specifies a scalar product
\begin{equation}\label{integrationonmanifold}
(f,h) = \int \sqrt{g} f h,
\end{equation}
and the adjoint of an operator $\mathcal{L}$ is thus given by
\begin{equation}
(f,\mathcal{L}^{\dagger}\left[h\right]) = (\mathcal{L}\left[ f\right], h) = \int \sqrt{g} h \mathcal{L}\left[f\right].
\end{equation}
These considerations are immediately extended to square integrable functions on the product manifold 
$\mathbb{R}^n\otimes \mathcal{M}$ which has the flat Euclidean metric on $\mathbb{R}^n$ and the metric $g$ on $\mathcal{M}$.
Let $\mathbb{R}^n$ be divided in sets $B^{(n)}_j$ and let $\mathbb{R}^n\otimes \mathcal{M}$ be divided in the sets $B^{(n\times m)}_{ji}=B^{(n)}_j \otimes B_i$.
For the generator of diffusions in (\ref{generator of diffusions}) the adjoint can be computed by integration by parts in each set $B^{(n\times m)}_{ji}$. It is a fundamental
fact that the boundary conditions then cancel between neighbouring sets and 
\begin{widetext}
\begin{equation}\label{OperatorMdagger}
 \mathcal{L}^{\dagger}=\frac{1}{\sqrt{g}}\left[-\partial_{i}\sqrt{g}A_{1}^i-\frac{\partial}{\partial  \alpha^a}\sqrt{g}A_{2}^a +\frac{1}{2}\partial^{2}_{ij}\sqrt{g}b_{11}^{ij}+\frac{1}{2}\frac{\partial^{2}}{\partial \alpha^{a} \partial \alpha^{b}}\sqrt{g}b_{22}^{ab} +\partial_{i}\frac{\partial}{\partial \alpha^{a}}\sqrt{g}b_{12}^{ai} \right]. 
\end{equation}
\end{widetext}
By comparing (\ref{OperatorMdagger}) and (\ref{Fokker-Planck-operator}) we can write the Fokker-Planck operator as
\begin{equation}
\label{eq:FP-Ldagger}
\mathcal{F}\left[P\right] = \sqrt{g} \mathcal{L}^{\dagger}\left[\frac{P}{\sqrt{g}}\right].
\end{equation}
For the following it is convenient to write 
\begin{equation}
\label{eq:P-hat-definition}
P = \hat{P} \sqrt{g},
\end{equation} 
where $\hat{P}$ is a scalar function. 

\subsection{Closed solutions for the stationary state}
\label{s:closed-solutions}
The stationary state of the Fokker-Planck equation in ordinary space can be found in closed
form if either the diffusion process satisfies detailed balance, or when the drift field is
incompressible. In this subsection we state the generalization of these results to diffusions
on a compact manifold. 
To streamline the notation we use here only one $d$-dimensional
drift vector $A$ with components $A^{\mu}$ and one $d\times d$-dimensional 
diffusion matrix $b$ with components $b^{\mu\nu}$. 
Taking the manifold to be ${T}^n\otimes \mathcal{M}$ where  ${T}^n$
is the $n$-dimensional torus ($d=m+n$), these stationary states
can then be used as zeroth-order probabilities in the 
multi-scale analysis in Section~\ref{s:multi-scale-internal}
and following. 

The starting point is the observation that the H\"anggi-Klimontovich drift
\begin{equation}
\label{eq:HK}
D^{\mu} = A^{\mu}-\frac{1}{2\sqrt{g}}\partial_{\nu}\left(\sqrt{g}b^{\mu\nu}\right)
\end{equation}
transforms as a covariant vector. For completeness a demonstration is given in Appendix.
If for some scalar function $V$ and in some coordinate system we have
\begin{equation}
\label{eq:gradient-field}
D^{\mu} = - \frac{1}{2} b^{\mu\nu} \partial_{\nu} V, 
\end{equation}
then this must therefore be true in every coordinate system. 
The flux-less stationary states of (\ref{Fokker-Planck-operator}) are determined by 
\begin{equation}
\label{eq:flux-less-1}
\frac{1}{2}b^{\mu\nu}\left(\partial_{\nu}P\right) =  \left(A^{\mu} - \frac{1}{2}\partial_{\nu}\left(b^{\mu\nu}\right)\right)P.
\end{equation}
Using the scalar $\hat{P}=P/\sqrt{g}$ we have instead
\begin{equation}
\label{eq:flux-less-2}
\frac{1}{2}b^{\mu\nu}\left(\partial_{\nu}\log \hat{P}\right) =  \left(A^{\mu} - \frac{1}{2}\partial_{\nu}\left(b^{\mu\nu}\right)
-\frac{1}{2}b^{\mu\nu}\frac{\partial_{\nu}\sqrt{g}}{\sqrt{g}}\right) = D^{\mu}.
\end{equation}
Therefore, if (\ref{eq:gradient-field}) is true the stationary state is given by
\begin{equation}
\label{eq:Gibbs}
P^* = \frac{1}{\mathcal{N}}\sqrt{g}e^{-V},
\end{equation}
where $\mathcal{N}$ is a normalization constant.
The mean drift with respect to this measure must vanish:
\begin{eqnarray}
\label{eq:zero-drift}
V^{\mu} &=& \frac{1}{\mathcal{N}}\int \sqrt{g}e^{-V}  A^{\mu} \nonumber \\
       &=& \frac{1}{\mathcal{N}}\int \sqrt{g}e^{-V} D^{\mu} - \frac{1}{\mathcal{N}}\int \sqrt{g}e^{-V} D^{\mu} = 0. 
\end{eqnarray}
Therefore we can only use (or hope to use) this solution method when the average drift in the internal
states vanishes, which for a rotating Brownian particles means that it cannot have a mean rotation around an axis.
On the other hand, for motion in space we can go to a comoving frame where the average drift is zero, and
then compute the effective diffusion in that frame. 
 
The other class of closed-form solutions is obtained by noting that 
\begin{equation}
\label{eq:covariant}
c=\partial_a\left( D^a\right) + D^a \frac{\partial_a\sqrt{g}}{\sqrt{g}} 
\end{equation}
is a scalar. The condition $c=0$ is thus the natural generalization of incompressibility, $\partial_{\mu}A^{\mu}=0$ in local flat coordinates.
Consider then
\begin{equation}
\label{eq:covariant}
S=\frac{1}{2} \int\sqrt{g} \left(\hat{P}-\frac{1}{\mathcal{N}}\right)^2,
\end{equation}
where $\int\sqrt{g} \hat{P}=\int\sqrt{g}\frac{1}{\mathcal{N}} =1$.
Using (\ref{OperatorMdagger}) we have
\begin{eqnarray}
\label{eq:covariant-time-differentiation}
\partial_t S &=& \int\sqrt{g} \left(\hat{P}-\frac{1}{\mathcal{N}}\right)\partial_t \hat{P} \nonumber \\
            &=& 2 \int\sqrt{g} c \hat{P}^2-\frac{1}{2} \int\sqrt{g} b^{ab}\partial_a\hat{P}\partial_b\hat{P}.
\end{eqnarray}
If $c=0$ the solutions must therefore relax to $\hat{P}=\frac{1}{\mathcal{N}}$. 

\section{Multi-scale with internal states}
\label{s:multi-scale-internal}
In a pioneering contribution published almost two decades ago
Vergassola and Avellaneda showed that scalar transport in a velocity field that varies on a small scale
leads to advection and diffusion on large scales with transport coefficients that can be computed
in a hierarchy of solvability conditions and solutions~\cite{VergassolaAvellaneda}; further results
in the same direction were later obtained in \cite{MazzinoMusacchioVulpiani05} and \cite{MartinsMazzinoPMG12}.
We will here apply the same method to the analysis of the Fokker-Planck equation on a product manifold $\mathbb{R}^n\otimes \mathcal{M}$ written
\begin{equation}
\label{eq:FokkerPlanck}
\partial_t \hat{P} =  \mathcal{L}^{\dagger}\hat{P},
\end{equation}
where $\mathcal{L}^{\dagger}$ is defined in (\ref{OperatorMdagger}) and where the auxiliary scalar function 
$\hat{P}$ is defined in (\ref{eq:P-hat-definition}).
From a technical point of view the material in this Section is not very new, except that we keep track of the manifold $\mathcal{M}$ of internal
states which adds to the notational complications; readers
familiar with the multi-scale formalism may want to skip to the following Section~\ref{s:internal-symmetry}. 

The first step is to identify a characteristic time $t_{\mathcal{M}}$ in which diffusion spreads out probability mass on the manifold $\mathcal{M}$.
By order of magnitude $t_{\mathcal{M}}\sim D_{\mathcal{M}}^2/b$ where $D_{\mathcal{M}}$ is the diameter of $\mathcal{M}$ and $b$ is a characteristic size of
the diffusion coefficients $b_{22}$ in (\ref{OperatorMdagger}). The second step is to identify a characteristic spatial scale
$L_{\mathcal{M}}$ which the diffusion process will reach during time $t_{\mathcal{M}}$. For the case of diffusion of the position and orientation of a
three-dimensional body in space and on $SO(3)$, and for physically reasonable diffusion coefficients, $L_{\mathcal{M}}$ is on the order of the 
radius of the body~\cite{5Authors}. We will from now on assume that scales of time and space and distances on the manifold have been chosen
such that $t_{\mathcal{M}}$, $L_{\mathcal{M}}$ and $D_{\mathcal{M}}$ are all of order one, meaning that all the diffusion coefficients
$b_{11}$, $b_{22}$ and $b_{12}$ in (\ref{OperatorMdagger}) are also of order one.
We will further assume that in the same coordinates the drift coefficients $A_{1}^i$ and $A_{2}^a$ are also of order one.
If in fact $A_{1}^i$ and $A_{2}^a$ would be smaller this would mean that the drift is relatively small (a special case of what will be considered below)
while if  $A_{1}^i$ and $A_{2}^a$ would be larger then diffusion would not be the fastest process on the scale of the internal states, and the starting point of 
the analysis should be different.
The third step is to identify a larger spatial scale $L$ and a small dimension-less ratio $\epsilon=\frac{L_{\mathcal{M}}}{L}$. We seek an effective description
of the motion in space on scale $L$. To do so we assume there are processes on time scale $t_{\mathcal{M}}$ which act to spread out probability mass
over $\mathcal{M}$ and distances $L_{\mathcal{M}}$ in space, processes on time scale  $\epsilon^{-1}t_{\mathcal{M}}$ where the probability mass is advected over 
length scale $L$ and processes on time scale  $\epsilon^{-2}t_{\mathcal{M}}$ where the probability mass diffuses relative distance $L$. 

The multi-scale step proper is let physical space be represented by two variables $\tilde{x}$ on scale $L_{\mathcal{M}}$ and $X$ on scale $L$
and similarly time by  $\tilde{t}$ on scale $t_{\mathcal{M}}$, $\tau$ on scale $\epsilon^{-1}t_{\mathcal{M}}$ and $\theta$ on scale  $\epsilon^{-2}t_{\mathcal{M}}$.
Derivatives with respect to physical time and space are then represented as
\begin{equation}
 \begin{split}
  & \frac{\partial}{\partial t} \to \partial_{\tilde{t}}+\epsilon \partial_{\tau}+\epsilon^{2}\partial_{\theta},\\
  & \frac{\partial}{\partial x^{i}} \to \partial_{i}+\epsilon \nabla_{i}, 
  \end{split}
\end{equation}
where $\partial_{i}$ stands for $\frac{\partial}{\partial \tilde{x}^i}$ and  $\nabla_{i}$ for $\frac{\partial}{\partial X^i}$ and 
$\epsilon$ is the small parameter.
The probability density function is expressed in local coordinates and expanded as
\begin{widetext}
 \begin{eqnarray}
\label{eq:P-expansion}
P&=&P^{(0)}(\tilde{x},X,\alpha,\tilde{t},\tau,\theta)+\epsilon P^{(1)}(\tilde{x},X,\alpha,\tilde{t},\tau,\theta)+\epsilon^2 P^{(2)}(\tilde{x},X,\alpha,\tilde{t},\tau,\theta) +\ldots\nonumber \\
&=&\sqrt{g}\left(\hat{P}^{(0)}(\tilde{x},X,\alpha,\tilde{t},\tau,\theta)+\epsilon \hat{P}^{(1)}(\tilde{x},X,\alpha,\tilde{t},\tau,\theta)+\epsilon^2 \hat{P}^{(2)}(\tilde{x},X,\alpha,\tilde{t},\tau,\theta) +\ldots\right)
\end{eqnarray}
\end{widetext}
and (\ref{eq:FokkerPlanck}) 
is solved order by order in $\epsilon$.
All functions $P^{(0)}, P^{(1)}, P^{(2)},\ldots$ are assumed periodic in the small-scale variable $\tilde{x}$ with period $L_{\mathcal{M}}$.

An important role is now played by the part of the operator $\mathcal{L^{\dagger}}$ in (\ref{OperatorMdagger}) which is of leading order in $\epsilon$.
In local coordinates $(\tilde{x},\alpha)$ it is written in the same way as  (\ref{OperatorMdagger}), \textit{i.e.} 
\begin{widetext}
\begin{equation}\label{OperatorLdagger}
 \mathcal{L}_0^{\dagger}=\frac{1}{\sqrt{g}}\left[-\partial_{i}\sqrt{g}A_{1}^i-\frac{\partial}{\partial  \alpha^a}\sqrt{g}A_{2}^a+\frac{1}{2}\partial^{2}_{ij}\sqrt{g}b_{11}^{ij}+\frac{1}{2}\frac{\partial^{2}}{\partial \alpha^{a} \partial \alpha^{b}}\sqrt{g}b_{22}^{ab} +\partial_{i}\frac{\partial}{\partial \alpha^{a}}\sqrt{g}b_{12}^{ia} \right],
\end{equation}
\end{widetext}
but is an elliptic partial differential operator not on  $R^n\otimes \mathcal{M}$, but on the compact
manifold $T^n\otimes \mathcal{M}$, where  $T^n$ is the $n$-dimensional torus with radii $L_{\mathcal{M}}$.
The coefficients of $\mathcal{L}_{0}^{\dagger}$ depend in principle on $\tilde{x}$, $X$ and $\alpha$.
We now appeal to standard results on spectra of elliptic operators on compact spaces;  $\mathcal{L}_0^{\dagger}$ should have a discrete spectrum  
\begin{equation}\label{spectrum}
\hbox{Spec}(\mathcal{L}_0)=\{\lambda_0, \lambda_1,\ldots\},
\end{equation}
and $\lambda_0=0$ is a unique zero mode. Physically this corresponds 
the solutions of (\ref{eq:FokkerPlanck}) relaxing for long time to a unique stationary state; as total probability mass
is conserved the corresponding eigenvalue must then be zero.
The operator $\mathcal{L}_0^{\dagger}$ and its adjoint $\mathcal{L}_0$ 
have the same eigenvalues but not necessarily the same eigenfunctions.
Let these be given as
\begin{equation}\label{eigprob}
\begin{split}
&\mathcal{L}_0n_l=\lambda_l n_l;\\
&\mathcal{L}_0^{\dagger}m_l=\lambda_l m_l,
\end{split}
\end{equation}
where $(n_l,m_k)=\mathbf{1}_{lk}$. Since in $\mathcal{L}_0$ all derivatives stand to the right of all variable-dependent coefficients, $n_0$ must 
be a constant. We can choose that to be $1$; the orthogonality condition for $m_0$ and $n_0$ then gives
a normalization of $m_0$:
\begin{equation}
\label{eq:m_0-normalization}
\int_{T^n\otimes \mathcal{M}}\sqrt{g}m_{0} = 1.
\end{equation}
The scalar density $\sqrt{g}m_{0}$ is then a normalized probability density on ${T^n\otimes \mathcal{M}}$.

The multi-scale is now posed by the following hierarchy:
\begin{widetext}
\begin{equation}\label{Hierarchyofequation}
\begin{split}
 & \mathcal{L}_0^{\dagger}\hat{P}^{(0)}=0;\\
 & \mathcal{L}_0^{\dagger}\hat{P}^{(1)}=\partial_{\tau}\hat{P}^{(0)}+\frac{1}{\sqrt{g}}\nabla_{i}[\sqrt{g}A_{1}^i \hat{P}^{(0)}]-\frac{1}{\sqrt{g}}\partial_{i}\nabla_{j}[\sqrt{g} b_{11}^{ij}\hat{P}^{(0)}]-\frac{1}{\sqrt{g}}\frac{\partial}{\partial \alpha^{a}}\nabla_{i}[\sqrt{g} b_{12}^{ia}\hat{P}^{(0)}];\\
 & \mathcal{L}_0^{\dagger}\hat{P}^{(2)}=\partial_{\tau}\hat{P}^{(1)}+\frac{1}{\sqrt{g}}\nabla_{i}[\sqrt{g}A_{1}^i\hat{P}^{(1)}]-\frac{1}{\sqrt{g}}\partial_{i}\nabla_{j}[\sqrt{g} b_{11}^{ij}\hat{P}^{(1)}]-\frac{1}{\sqrt{g}}\frac{\partial}{\partial \alpha^{a}}\nabla_{i}[\sqrt{g} b_{12}^{ia}\hat{P}^{(1)}]+\\
 &\quad  \partial_{\theta}\hat{P}^{(0)}-\frac{1}{2}\frac{1}{\sqrt{g}}\nabla_{i}\nabla_{j}[\sqrt{g}b_{11}^{ij}\hat{P}^{(0)}].
\end{split}
\end{equation}
\end{widetext}
All equations are solved in the relaxation limit for the fast time $\tilde{t}$. 
We note that by assumption $\sqrt{g}$ only depends on the internal coordinates, hence it commutes with $\nabla_i$, the derivative on the large scale.

\subsection{Dependence of $P^{(0)}$ on $\tilde{x}$ and internal states}
The solution to zeroth order is
\begin{equation}
 \hat{P}^{(0)}(\vec{\tilde{x}},\vec{\alpha},\vec{X}, \tau, \theta) = m_{0}(\vec{\tilde{x}},\vec{\alpha};\tau,\theta,X) C^{(0)}(\vec{X},\tau, \theta),
\end{equation}
where $m_{0}(\vec{\tilde{x}},\vec{\alpha};\tau,\theta,X)$ is the zero mode of $\mathcal{L}_0^{\dagger}$ with normalization 
(\ref{eq:m_0-normalization}). Note that $m_{0}$ depends parametrically on $\tau$, $\theta$ and $X$ because the coefficients of 
$\mathcal{L}_{0}^{\dagger}$ may depend on all these variables.
$C^{(0)}$ is on the other hand a (so far undetermined) proportionality coefficient.
The zeroth other probability $P^{(0)}$ is then $\sqrt{g} m_{0} C^{(0)}$.
Except for the special cases discussed above in Subsection~\ref{s:closed-solutions}
solving for $m_{0}$ is in general not straight-forward
and requires numerical methods. For the general discussion in this section it is however enough
to assume that it exists, and that it depends smoothly on  $\tilde{x}$, $\alpha$ and on $(\tau,\theta,X)$.   

\subsection{Solvability conditions and solution to order $\epsilon$}
Generally we must have 
\begin{equation}
\mathcal{L}_0^{\dagger}\hat{P}^{(1)}\in \text{Im}(\mathcal{L}_0^{\dagger})=\text{Span}\{m_1, m_2,\ldots\},
\end{equation}
where the zero mode $m_0$ does not appear on the right-hand side. By Fredholm alternative we then have $(\mathcal{L}_0^{\dagger}\hat{P}^{(1)}, n_0)=0$
and using the second line of (\ref{Hierarchyofequation}),  and the solution $P^{(0)}$ obtained above
we have a \textit{solvability condition} (two terms vanish because the are gradients with respect to the small scales that are integrated over):
\begin{equation}
(n_0, \frac{\partial}{\partial \tau}(m_0C^{(0)})) + \nabla_i(n_0, A_1^i m_0C^{(0)}).
\label{equation1}
\end{equation}
Since $C^{(0)}$ does not depend on the small scales, the first term in (\ref{equation1}), 
$(n_0, \frac{\partial}{\partial \tau}(m_0C^{(0)}))$, is equal to $C^{(0)}(n_0, \frac{\partial}{\partial \tau}(m_0))+ \frac{\partial}{\partial \tau}(C^{(0)})(n_0,m_0)$.
On the other hand, by orthogonality  $(n_0, m_0)=1$ for any values of
$(\tau,\theta, X)$ and $n_0$ has been chosen independent of $\tau$. Therefore $(n_0, \frac{\partial}{\partial \tau}(m_0))=0$
and (\ref{equation1}) simplifies to

\begin{equation}\label{ballisticequation}
 \partial_{\tau}C^{(0)}(\vec{X},\tau, \theta)+\nabla_{i}[C^{(0)}(\vec{X},\tau, \theta)(n_0,A_{1}^im_0)]=0.
\end{equation}

Equation (\ref{ballisticequation}) describes advective motion with effective drift velocity
\begin{equation}
V^i(\tau,\theta,X)=(n_0,A_1^i m_0)=\int_{T^n\otimes \mathcal{M}}\sqrt{g}\,A_{1}^i  m_0 .
\label{eq:ballistic-velocity}
\end{equation}
The above is a straightforward generalization of the result given in Eqs~17 and~18 in~\cite{VergassolaAvellaneda}.

To \textit{solve} for $P^{(1)}$ we first write
\begin{widetext}
\begin{equation}
\mathcal{L}_0^{\dagger}\hat{P}^{(1)}=m_0\frac{\partial C^{(0)}(\vec{X},\tau, \theta)}{\partial \tau}+C^{(0)}(\vec{X},\tau, \theta)\frac{\partial m_0}{\partial \tau}+
\nabla_i [A_1^i  C^{(0)}(\vec{X},\tau, \theta)m_0] +\tilde{P}^{(1)},
\label{eq:L-to-solve-first-order}
\end{equation}
where the remainder term is
\begin{equation}
\tilde{P}^{(1)}=-\frac{1}{\sqrt{g}}\partial_{i}\nabla_{j}[b_{11}^{ij}P^{(0)}]-\frac{1}{\sqrt{g}}\frac{\partial}{\partial \alpha^{a}}\nabla_{i}[b_{12}^{ia}P^{(0)}] \in \text{Im}(\mathcal{L}_0^{\dagger}).
\end{equation}
\end{widetext}
Using the solvability condition the first term in (\ref{eq:L-to-solve-first-order})
can be re-written $-m_0\nabla_i[C^{(0)}V^i]$, and we have
\begin{widetext}
\begin{equation}
%\begin{split}
\mathcal{L}_0^{\dagger}\hat{P}^{(1)}=C^{(0)}\left(\frac{\partial m_0}{\partial \tau}+V^i\nabla_i m_0\right)+\nabla_i [C^{(0)}\left(A_1^i-V^i\right)m_0]+\tilde{P}^{(1)}.
\label{eq:L-to-solve-first-order-v2}
%\end{split}
\end{equation}
\end{widetext}
By the same argument as given above $(n_0,\nabla_i m_0)=0$, and all 
terms on the right hand side of (\ref{eq:L-to-solve-first-order-v2})
are therefore in $\text{Im}(\mathcal{L}_0^{\dagger})$. The inverse $(\mathcal{L}_0^{\dagger})^{-1}$ is an integral operator on functions on $T^n\otimes \mathcal{M}$
defined on all functions orthogonal to $n_0$, and 
\begin{widetext}
\begin{eqnarray}
\hat{P}^{(1)}&=&C^{(0)}\left((\mathcal{L}_0^{\dagger})^{-1}\frac{\partial m_0}{\partial \tau}+V^i(\mathcal{L}_0^{\dagger})^{-1} \nabla_i m_0\right)+
[\nabla_i C^{(0)}](\mathcal{L}_0^{\dagger})^{-1}[\left(A_1^i-V^i\right)m_0] \nonumber \\
&& +C^{(0)}(\mathcal{L}_0^{\dagger})^{-1} \nabla_i [\left(A_1^i-V^i\right)m_0] +(\mathcal{L}_0^{\dagger})^{-1} \tilde{P}^{(1)} + \hat{P}^{(1),\text{hom}},
\label{eq:solution-first-order}
\end{eqnarray}
\end{widetext}
where $\hat{P}^{(1),\text{hom}}$ is the homogenous solution which can be written $C^{(1)}m_0$, where $C^{(1)}$ is another proportionality.
The way in which $(\mathcal{L}_0^{\dagger})^{-1}$ does not commute with $\nabla_i$ is illustrated by second and third terms in (\ref{eq:solution-first-order}).
Even though $(\mathcal{L}_0^{\dagger})^{-1}$ acts on functions on $T^n\otimes \mathcal{M}$ its coefficients depend (by assumption) on the slow variables. 
To bring $(\mathcal{L}_0^{\dagger})^{-1}$ completely inside the action of $\nabla_i$ we would have to write these two terms as
\begin{widetext}
\begin{equation}
\nabla_i\left[ C^{(0)}(\mathcal{L}_0^{\dagger})^{-1}[\left(A_1^i-V^i\right)m_0]\right]+C^{(0)}  (\mathcal{L}_0^{\dagger})^{-1} [\nabla_i\mathcal{L}_0^{\dagger}] (\mathcal{L}_0^{\dagger})^{-1} \left[\left(A_1^i-V^i\right)m_0\right],
\nonumber
\end{equation}
\end{widetext}
where $(\mathcal{L}_0^{\dagger})^{-1} (\nabla_i\mathcal{L}_0^{\dagger}) (\mathcal{L}_0^{\dagger})^{-1}$ is another operator. 

\subsection{Solvability conditions to order $\epsilon^2$}
The effective diffusion on large scales if determined by the solvability condition on second order \textit{i.e.}
\begin{equation}
\mathcal{L}_0^{\dagger}\hat{P}^{(2)} \in \text{Im}(\mathcal{L}_0^{\dagger}).
\end{equation}
Considering the right hand side of the third line of (\ref{Hierarchyofequation})
this means
\begin{widetext}
\begin{equation}
\label{equation2}
\partial_{\tau}[(n_0,\hat{P}^{(1)})] + \nabla_i[(n_0,A_1^i \hat{P}^{(1)})] + \partial_{\theta}C^{(0)} - \frac{1}{2} \nabla_i \nabla_j [(n_0,b_{11}^{ij}m_0)C^{(0)}]=0,
\end{equation}
\end{widetext}
where $\hat{P}^{(1)}$ is given by (\ref{eq:solution-first-order}). 
The last term gives a first effective diffusion term with effective diffusion coefficient matrix
\begin{equation}
\label{eq:first-effective-diffusion}
\kappa^{ij}_{(1)}=\frac{1}{2} (n_0,b_{11}^{ij}m_0).
\end{equation}
The homogeneous part $\hat{P}^{(1),\text{hom}}$ enters in the first two terms in (\ref{equation2})
as a continuity equation for the proportionality $C^{(1)}$ with the same effective drift velocity as in (\ref{eq:ballistic-velocity}),
while the other terms in $P^{(1)}$ can be written
\begin{equation}
\hat{P}^{(1)}- \hat{P}^{(1),\text{hom}} = [\nabla_kC^{(0)}]\chi^k+C^{(0)}\xi, 
\end{equation}
with two auxiliary functions $\chi^k$ (given below) and $\xi$.
Overall the second term in (\ref{equation2}) therefore contributes 
\begin{widetext}
\begin{equation}\label{generalequationeffdiff}
\nabla_i[(n_0,A_1^i \hat{P}^{(1)})] = \nabla_i\nabla_k\left[C^{(0)}(n_0,A_1^i \chi^k)\right] + \nabla_i\left[C^{(0)}(n_0,A_1^i \xi)-\nabla_k [(n_0,A_1^i \chi^k)]\right] 
\end{equation}
\end{widetext}
of which the first is a diffusion and the second is a higher-order advection with effective drift velocity 
$V^i_1=\left[(n_0,A_1^i \xi)-\nabla_k [(n_0,A_1^i \chi^k)]\right]$.
The second effective diffusion term therefore depends only on 
\begin{widetext}
\begin{equation}\label{GeneralAuxEq}
\chi^k=  (\mathcal{L}_0^{\dagger})^{-1}\left[(A_1^k - V^k)m_0 - \partial_{j} \left(b_{11}^{kj}m_0\right) - \frac{1}{\sqrt{g}}\frac{\partial}{\partial \alpha^{a}}\left(\sqrt{g} b_{12}^{ka}m_0\right)\right], 
\end{equation}
\end{widetext}
and we can also write
\begin{widetext}
\begin{equation}
\label{eq:second-effective-diffusion}
\kappa^{ik}_{(2)}=\left((\mathcal{L}_0)^{-1}\left[\left(V^i-A_1^i\right)n_0\right],(A_1^k - V^k)m_0 - \partial_{j} \left(b_{11}^{kj}m_0\right) - \frac{1}{\sqrt{g}}\frac{\partial}{\partial \alpha^{a}}\left(\sqrt{g} b_{12}^{ka}m_0\right)\right).
\end{equation}
\end{widetext}
Naturally only the components symmetric in interchanging indices $i$ and $k$ matter in the above; (\ref{eq:second-effective-diffusion}) is hence the generalization of Eq.~64 in~\cite{VergassolaAvellaneda}.

\subsection{Semi-analytic solutions for $\chi^k$}
\label{s:semi-analytic-solution}
For generalized potential motion discussed above, where $m_0=\frac{1}{\mathcal{N}_{(V)}}e^{-V}$ and where $\partial_a V = -2(b^{-1})_{ab}D^b$ where $D^b$ is the H\"anggi-Klimontovich drift, the calculation of the transport coefficient can be simplified. 
First recall that for consistency we must then be in a comoving frame where the drift velocity $V^k$ is zero.
Equation (\ref{GeneralAuxEq}) determining $\chi^k$ is then 
\begin{equation}
%\frac{1}{2\sqrt{g}}\partial_{\mu}e^{-V}\left(\sqrt{g}b^{\mu\nu}\partial_{\nu}(e^{V}\chi^k_{(V)})\right) = -A_1^k\frac{1}{\mathcal{N}_{(V)}}e^{-V}
\mathcal{L}_V^{\dagger}\chi^k_{(V)}= -A_1^k\frac{1}{\mathcal{N}_{(V)}}e^{-V},
 \label{eq:solution-first-order-2}
 \end{equation}
where $\mathcal{L}_V^{\dagger}[\cdot]=\frac{1}{2\sqrt{g}}\partial_{\mu}\left(e^{-V}\sqrt{g}b^{\mu\nu}\partial_{\nu}(e^{V}[\cdot])\right)$. 
%On the other hand the operator itself is then (having $\mu,\nu$ for all the directions on the small scales)
%\begin{equation}
% \label{eq:solution-first-order-3}
%\mathcal{L}_0^{\dagger}\chi^k_{(V)} = \frac{1}{2\sqrt{g}}e^{-V}\partial_{\mu}\left(\sqrt{g}b^{\mu\nu}\partial_{\nu}(e^{V}\chi^k_{(V)})\right)
%\end{equation}
Suppose now that for some drift field $A_2$ in the internal states, \textit{e.g} $A_2=0$ in local flat coordinates, we have found as solution the uniform 
measure $m_0=\frac{1}{\mathcal{N}_{(0)}}$. For that drift field there is an auxiliary field $\chi^k_{(0)}$ which satisfies
\begin{equation}
\mathcal{L}_0^{\dagger}\chi^k_{(0)} = -A_1^k\frac{1}{\mathcal{N}_{(0)}},
 \label{eq:solution-first-order-4}
\end{equation}
and the contribution to the effective diffusion coefficient (\ref{eq:second-effective-diffusion}) is
\begin{equation}
\label{eq:second-effective-diffusion-V-is-zero}
\kappa^{ik}_{(2)}(V=0)= -\frac{1}{\mathcal{N}_{(0)}} \int\sqrt{g} \chi^k_{(0)} A_1^i .
\end{equation}
%Comparing  (\ref{eq:solution-first-order-2}) and  (\ref{eq:solution-first-order-4}) it is however obvious that
A solution of (\ref{eq:solution-first-order-2}) can be found by:
\begin{equation}
\label{eq:solution-49}
%\chi^k_{(V)} = \frac{\mathcal{N}_{(0)}}{\mathcal{N}_{(V)}}e^{-V} \chi^k_{(0)}
\chi^k_{(V)} = - (\mathcal{L}_V^{\dagger})^{-1}\left[ A_1^k\frac{1}{\mathcal{N}_{(V)}}e^{-V}\right],
\end{equation}
and for general potential motion we thus have
\begin{equation}
\label{eq:second-effective-diffusion-V}
\kappa^{ik}_{(2)}(V)= - \frac{1}{\mathcal{N}_{(V)}} \int\sqrt{g} \chi^k_{(V)} A_1^i.
\end{equation}
%We therefore have to average \textit{the same} quantities for every potential $V$ but the weighting
%in the average is \textit{different} for different $V$.}
We, therefore, must solve equation (\ref{eq:solution-first-order-2}) for every potential and
average on the manifold.

\subsection{Collapsing equation on orders $\epsilon$ and $\epsilon^2$}
\label{s:collapsing-orders} 
To bring out one final equation we introduce the adjusted 
advective velocity $\tilde{V}^i=V^i +\epsilon V^i_1$
and the adjusted zero-order proportionality 
\begin{equation}
\tilde{C}=C^{(0)}+\epsilon C^{(1)},
\end{equation}
and bring back the original variables in time and space.
By above we then have
\begin{widetext}
\begin{equation}
\label{eq:Fokker-Planck-large-scales}
\partial_{t}\tilde{C}=-\partial_i\left[\tilde{V}^i \tilde{C}\right] + \partial_i\partial_k\left[( \kappa^{ik}_{(1)}+\kappa^{ik}_{(2)})\tilde{C}\right], 
\end{equation}
\end{widetext}
where all terms are of order $(\epsilon)^2$ (diffusive time scale), and where the first correction is of order $(\epsilon)^3$.

\section{Multi-scale with internal symmetry}
\label{s:internal-symmetry}
In this section we apply the general results of the previous section to the physical setting where the 
system of stochastic differential equations (\ref{Overdampedequations}) are overdamped equations of motion
and where the manifold describes a symmetry of the motion in space.
We then assume a drift field $\vec{F}$ in the spatial directions which we call force, and a drift field $\vec{M}$ in the 
tangent space of the manifold which we call torque.

\subsection{Overdamped motion}
We assume that $\vec{F}$ only depends on the large scale $\vec{X}$ but not on the internal state nor on the small scale $\vec{\tilde{x}}$, while $\vec{M}$ may depend on both 
the internal state and the large scale $\vec{X}$.
Further assume a friction matrix 
\begin{equation}
\Gamma=\left(\begin{matrix}     \gamma_{11} & \gamma_{12} \\ \gamma_{21} & \gamma_{22}\end{matrix}\right), \nonumber
\end{equation}
which has to have only real positive eigenvalues such that the underdamped motion
on a faster time scale is purely relaxational.
In the examples discussed below of motion on $\mathbb{R}^3$ or $\mathbb{C}^2$ $\Gamma$ will be real symmetric or Hermitian.
$\Gamma$ therefore has a square root in the sense that $\Gamma=\sigma\sigma^T$  
\begin{equation}
\sigma=\left(\begin{matrix}     \sigma_{11} & \sigma_{12} \\ \sigma_{21} & \sigma_{22}\end{matrix}\right) \nonumber
\end{equation} 
such that Eq.~(\ref{Overdampedequations}) can be written
\begin{widetext}
\begin{equation}
\label{Overdampedequations-physical}
\begin{split}
& {\gamma_{11}}^i_j dx^j + {\gamma_{12}}^i_b d\alpha^b = F^idt + \sqrt{2k_BT}\left[(\sigma_{11})^i_j dW^j + (\sigma_{12})^i_b d{W'}^b\right], \\   
& {\gamma_{21}}^a_j dx^j + {\gamma_{22}}^a_b d\alpha^b = M^adt + \sqrt{2k_BT}\left[(\sigma_{21})^a_j dW^j + (\sigma_{22})^a_b d{W'}^b\right],
\end{split}
\end{equation} 
\end{widetext}
where $T$ is the temperature. 
The inverse of the friction matrix is a symmetric mobility matrix
\begin{equation}
\Gamma^{-1}=\left(\begin{matrix}     \mu_{11} & \mu_{12} \\ \mu_{21} & \mu_{22} \end{matrix}\right), \nonumber
\end{equation}
and the drift fields in (\ref{Overdampedequations}) are given by 
\begin{eqnarray}
\label{eq:drift-physical}
A_1^i &=& {\mu_{11}}^i_j F^j +  {\mu_{12}}^i_a M^a, \\
\label{eq:rotation-physical}
A_2^i &=& {\mu_{21}}^a_j F^j +  {\mu_{22}}^a_b M^a, 
\end{eqnarray}
while the diffusion amplitudes in  (\ref{Overdampedequations}) are
\begin{equation}
\label{eq:B-overdamped}
B =  \sqrt{2k_BT} \Gamma^{-1}\sigma.
\end{equation} 
We will for the rest of this section assume that all It\^o terms and other correction terms from
spatial variation of temperature and friction matrix have been included in the physical force and torque as the need may be.
The diffusion terms in the Fokker-Planck operator (\ref{OperatorMdagger}) 
follow from (\ref{eq:B-overdamped}) and the expressions listed below Eq.~(\ref{generator of diffusions}).
We will assume that
the program of the previous section can be carried out, that we can compute $m_0$, $P^{(1)}$ etc. as needed.

Before introducing symmetry, let us remark that the advection velocity (\ref{eq:ballistic-velocity-overdamped}) can now be written as
\begin{equation}
V^i(\tau,\theta,X)=(n_0,A_1^i m_0)= \overline{\mu}^i_j F^j +  \overline{V}^i
\label{eq:ballistic-velocity-overdamped}
\end{equation}
with an effective mobility 
\begin{equation}
\overline{\mu}^i_j = \int\sqrt{g} {\mu_{11}}^i_j m_0,
\label{eq:effective-mobility}
\end{equation}
and a drift generated by torsion and cross-mobility
\begin{equation}
\overline{V}^i = \int\sqrt{g} {\mu_{12}}^i_a M^a m_0.
\label{eq:effective-drift}
\end{equation}
If two bodies experience opposite torques but are otherwise equivalent they would hence typically migrate at different speeds when the 
cross-mobility $\mu_{12}$ is non-zero.

\subsection{Overdamped motion with a symmetry}
\label{s:overdamped-symmetry}
Let us now assume that the internal states are a symmetry group of the motion in $n$-dimensional space. 
Abstractly defined, a \textit{group} $G$ is a collection of elements $g$ with the following properties:
\begin{itemize}
\item there is a rule for multiplying any two elements, and their product $g_1g_2$ is also an element of $G$; this rule for multiplication is \textit{associative}, so for any three element of $(g_1g_2)g_3=g_1(g_2g_3)$;
\item there is an identity element of $G$, say $e$, such that $eg\,=\,ge\,=\,g$ for any element $g$;
\item for every element $g$, there is a unique inverse element $g^{-1}$ such that $gg^{-1}\,=\,g^{-1}g\,=\,e$.
\end{itemize}  
That the manifold $\mathcal{M}$ is a symmetry group of $\mathbb{R}^n$ means that there is a map from points $g$ on $\mathcal{M}$ to operators $R(g)$
acting on $\mathbb{R}^n$ which are a faithful representation of the group \textit{i.e.}     
\begin{itemize}
\item $R(g_1g_2)=R(g_1)R(g_2)$; 
\item $R(e)=\textbf{1}$, such that $R(eg)=R(ge)=R(g)$ for any element $g$;
\item $R(g^{-1})=(R(g))^{-1}$;
\end{itemize}  
All the products in the above are ordinary matrix products.
Let now $U_e$ be a patch of the manifold around the identity element $e$ and let $V_e=\psi_e(U_e)\in \mathbb{R}^m$ be local coordinates for that patch.
By the group action we can construct patches $U_g=g(U_e)$ around each point $g$, and we can therefore construct local coordinates for patch $U_g$ as  
$V_g=\psi_e(g^{-1}(U_g))$. 
From now on we will consider motions (\ref{Overdampedequations}) expressed in these local coordinates. 
Furthermore we will in this Section only consider patches that are very small so that the drift terms in (\ref{eq:drift-physical}) and
(\ref{eq:rotation-physical}) and the diffusion terms in (\ref{eq:B-overdamped}) are practically constant in each patch.
The values of the various quantities around $e$ will be indexed $B$ for ``body''.

That $\mathcal{M}$ is a symmetry group of the motion means that for any $g$ the overdamped motion described by
(\ref{Overdampedequations-physical}) should look the same if described in the equivalent local coordinates 
$V_g$ and in the transformed spatial coordinates 
\begin{equation}
{x'}^i = R^i_j(g) x^j.
\end{equation}
The deterministic and random forces in (\ref{Overdampedequations-physical}) are also be transformed in the same way, while
%\begin{widetext}
%\begin{eqnarray}
%&{F'}^i = R^i_j(g) F^j \\
%&\sqrt{2k_BT}\left[(\sigma'_{11})^i_j dW^j + (\sigma'_{12})^i_b d{W'}^b\right]= R^i_k \sqrt{2k_BT}\left[(\sigma'_{11})^k_j dW^j + (\sigma'_{12})^k_b d{W'}^%b\right]
%\end{eqnarray}
%\end{widetext}
the transformation properties of the Wiener noises is a matter of convention; here we will take them to be unchanged.
Clearly invariance of the motion in space only holds if the friction matrices transform as
\begin{eqnarray}
{\gamma}_{11}(g) &=& R(g) {\gamma^B}_{11} R^{-1}(g); \\
{\gamma}_{12}(g) &=& R(g) {\gamma^B}_{12} ;
\end{eqnarray}
and the spatial noise terms as
\begin{eqnarray}
{\sigma}_{11}(g) &=& R(g) {\sigma^B}_{11}; \\
{\sigma}_{12}(g) &=& R(g) {\sigma^B}_{12}. 
\end{eqnarray}
We have to assume that ${\gamma}_{21}$ and ${\gamma}_{22}$
transform as needed to preserve positivity; for a real symmetric friction matrix this means  
${\gamma}_{21}(g) = {\gamma}_{21} R^T(g)$ and  ${\sigma}_{21}(g) = {\sigma^B}_{21}R^T(g)$.

Although somewhat more general cases could be considered we will assume
that the friction and noise in the manifold directions are the same
at every point, \textit{i.e.}  ${\gamma}_{22}(g)={\gamma}^B_{22}$ and ${\sigma}_{22}(g)={\sigma}_{22}^B$.
The mobilities in (\ref{eq:drift-physical}) then transform as
\begin{eqnarray}
{\mu}_{11}(g) &=& R(g) {\mu^B}_{11} R^{-1}(g); \nonumber \\
{\mu}_{12}(g) &=& R(g) {\mu^B}_{12}; \nonumber 
\end{eqnarray}
and the effective mobility in (\ref{eq:effective-mobility}) is 
\begin{equation}
\overline{\mu}^i_j = \int\sqrt{g}  R^i_l ({\mu^B}_{11})^l_k {R^{-1}}^k_j m_0.
\label{eq:effective-mobility-2}
\end{equation}
As we will see below for the examples of $SO(3)$ and $SU(2)$ when $m_0$ is constant
this simplifies considerably and $\overline{\mu}^i_j$
then becomes proportional to the identity matrix.
The same result holds for the first (simple) component of the effective diffusivity in (\ref{eq:first-effective-diffusion}).

For invariance to be complete we would finally have to assume that the torque $M^a$ be the same at each point
of $\mathcal{M}$ (expressed in the local coordinates). Since physically the torque is (or could be) externally determined we will however 
relax this requirement, and the drift generated by cross-diffusion and torque is then 
\begin{equation}
\overline{V}^i = \int\sqrt{g}  R^i_l(g)  (\mu^B_{12})^l_a M^a m_0.
\label{eq:effective-drift-2}
\end{equation}
This can be non-zero if the variations of  $R^i_l(g)$ and $M^a m_0$ combine constructively, but would be zero if, for instance,
$M^a$ is constant and $m_0$ is uniform.

\section{Motion on $\mathbb{R}^3$ and rotations in the Lie group $SO(3)$}
\label{s:so3}
We will now make the previous section more concrete by considering translation and rotations in ordinary
three-dimensional space. 
The Lie group $SO(3)$ is the set of rotations about the origin of the three-dimensional Euclidean space $\mathbb{R}^3$ where the group product is composition of 
rotations. 
Physically the situation is described by two frames: the lab frame and the body frame, where everything that depends on the external world
(force $\vec{F}$ and perhaps partly torque $\vec{M}$) are simply expressed in the lab frame, and everything else is simply
expressed in the body frame. A rotation is a transformation that changes the body frame to the lab frame (active interpretation, left action of the group).
Rotations being linear transformations they can be represented as matrices once a basis of $\mathbb{R}^3$ has been chosen. Specifically, if we choose an orthonormal basis of $\mathbb{R}^3$, every proper rotation is described by an orthogonal $3\times3$ matrix with determinant one, which is the set $SO(3)$.
The following facts about $SO(3)$ are well known \textit{cf}~\cite{Choo_Book_Manifold}
\begin{itemize}
\item $SO(3)$ preserves the scalar product $(x,y)=\sum_i x^i y^i$. $SO(3)$ therefore preserves the Euclidean metric tensor $\hbox{diag}(1,1,1)$ which
is the Kronecker symbol $\delta_{ij}$. As a consequence, all spatial tensor indices can be lowered and raised at will. 
$SO(3)$ additionally preserves the three-dimensional volume element and the Levi-Civita tensor $\epsilon_{ijk}$.
\item $SO(3)$ is compact manifold of dimension $3$. 
\item The tangent space of $SO(3)$ at the identity element $e$ is the Lie algebra $so(3)$.
An element of the Lie algebra can be written $v=\sum_i \alpha_i L_i$ where $(\alpha_1,\alpha_2,\alpha_3)$ are three real parameters
and $(L_1,L_2,L_3)$ are three basis elements.  $so(3)$ is hence isomorphic to $\mathbb{R}^3$. 
\item The exponential map of the Lie algebra covers the group. 
\end{itemize} 
The last point means that every element $g\in SO(3)$ can be written $\exp(v)$ for some $v$.
This representation is of course not unique.
However, if the real parameters $\alpha$ lie sufficiently close to the origin then they 
determine a set $V_e\subset R^3$ such that the map $\alpha\to \exp(\sum_i \alpha_i L_i)$ is one-to-one.
The $\alpha$'s can therefore be used to construct a very convenient system of local coordinates
used in \cite{TaylorKriegmann} and earlier in~\cite{McKean1960}. 
In Appendix~\ref{LocalChartappendix} we summarize useful facts about this system of charts.

\subsection{No cross-diffusion and $m_0$ constant}
\label{s:first-simplification}
In this Section and in the next we will make successive assumptions to make the problem of computing 
large-scale advective-diffusive motion analytically solvable. The first of these assumptions is to assume that the cross-frictions
in (\ref{Overdampedequations-physical}) are absent. Our starting point is therefore the simpler set of overdamped equations
%\begin{widetext}
\begin{equation}
\label{Overdampedequations-physical-diagonal}
\begin{split}
& dx^i  = {\gamma^{-1}}^{i}_j F^jdt + \sqrt{2k_BT} (\gamma^{-\frac{1}{2}})^i_j dW^j , \\   
& d\alpha^b = {\eta^{-1}}^a_b M^bdt + \sqrt{2k_BT} (\eta^{-\frac{1}{2}})^a_b d{W'}^b, 
\end{split}
\end{equation} 
%\end{widetext}
where we have introduced $\gamma$ (spatial friction matrix) for $\gamma_{11}$ and  $\eta$ 
(rotation friction matrix) for $\gamma_{22}$. Note that for space (indices $i$ and $j$) 
we do not need to distinguish upper and lower indices, but for motion
on the group manifold (indices $a$ and $b$) we do.
We restate the assumptions already made on the dependence of the various quantities in (\ref{Overdampedequations-physical-diagonal}):
\begin{equation}\label{PhysicalQuantities}
\begin{split}
& T=T(\vec{X}, \tau, \theta); \\ 
& \gamma^{i}_j=\gamma^{i}_j(\vec{X}, \vec{\alpha}, \tau, \theta); \\
& \eta^{a}_b=\eta^{a}_b(\vec{X}, \tau, \theta); \\
& F^{i}=F^{i}(\vec{X},\tau,\theta); \\
& M^{a}=M^{a}(\vec{X},\vec{\alpha}, \tau, \theta).
\end{split}
\end{equation}
where $\gamma$, $\eta$ and $\vec{F}$ transform as discussed above in Section~\ref{s:overdamped-symmetry}.

A consequence of these assumptions is that the zero-order probability $P^{(0)} = \sqrt{g} C^{(0)} m_0$ does not depend on the small length scale
$\tilde{x}$. From now on we will additionally ignore the dependency of $T$, $\gamma$ and $\eta$ on $\tau$ and $\theta$ as they are not our concern here;
similarly to in Section~\ref{s:internal-symmetry}
above we have in (\ref{Overdampedequations-physical-diagonal})
also assumed that all It\^o terms and other correction terms from
spatial variation of temperature and friction matrix have been included in the physical force $\vec{F}$.

Second we assume that the dynamics is such that  $m_0$ is constant. As discussed above in Section~\ref{s:closed-solutions}, this is so when the drift
$(\eta^{-1})^a_bT^b$ is zero, and also when the generalized incompressibility condition in (\ref{eq:covariant}) is satisfied.
%As discussed in Appendix~\ref{LocalChartappendix} this reduces in the local charts to the simpler condition
%$(\eta^{-1})^a_b \partial_{\alpha^a} T^a=0$. 
One example when the latter holds is when $T^a$ is constant (a particle that tends to rotate around
an axis fixed in the body), another when the torque is constant in the lab frame.

We now discuss the effective mobility (\ref{eq:effective-mobility-2}) which we write as
\begin{equation}
\overline{\mu}_{ij} = \sum_{lk}\int\sqrt{g}  R_{il} (\gamma_B^{-1})_{lk} R^{-1}_{kj} m_0.
\label{eq:effective-mobility-3}
\end{equation}
where $R^{-1}_{kj}=R_{jk}$.
Consider the quadratic form $s(x,y)=\sum_{ij}\overline{\mu}_{ij} x_iy_j$ where  $x$ and  $y$ are two vectors.
By above this is
\begin{equation}
s(x,y) = \sum_{lk} \int\sqrt{g}  (\sum_i x_i R_{il}) (\gamma_B^{-1})_{lk}  (\sum_j y_i R_{jk}) m_0. \nonumber
\label{eq:effective-mobility-4}
\end{equation} 
As $m_0$ is constant the value of the integral is the same for every joint rotation of the two vectors; $s(Rx,Ry) = s(x,y)$.
Since the quadratic form invariant by $SO(3)$ is the Kronecker delta 
and since $\hbox{Tr} R (\gamma_B^{-1}) R^{-1}= \hbox{Tr} (\gamma_B^{-1})$ we therefore have
\begin{equation}
\overline{\mu}_{ij} = \frac{\text{Tr}(\gamma^{-1})}{3} \mathbf{1}_{ij},
\end{equation}
and the advective velocity is
\begin{equation}\label{ballisticvelocitySO(3)}
V_{i}=\frac{\text{Tr}(\gamma^{-1})}{3} F_{i}.
 \end{equation}
The same reasoning gives the first component of the effective diffusion tensor 
obtained in (\ref{eq:first-effective-diffusion}) as
\begin{equation}
\label{eq:recall}
\kappa^{ij}_{(1)}= k_BT \frac{\text{Tr}(\gamma^{-1})}{3} \hbox{1}^{ij}.
\end{equation}
The invariant method used here is described in is more detail in Appendix~\ref{InvariantMethodAppendix}.

\subsection{No torque and $\eta$ diagonal}
\label{s:no-torque-eta-diagonal}
To proceed further we need to solve for the first order probability $P^{(1)}$, and to do so in closed
form we assume that there is no torque. We will also assume that the angular friction matrix
$\eta$ is diagonal which amounts to a choice of basis for the body frame. Note that we will not assume
that $\gamma$ is diagonalizable in the same basis, a simplifying assumption made in~\cite{5Authors} and~\cite{Brenner1981}. 

Let us first recall that the goal is to compute the second component of the effective diffusion tensor
obtained in (\ref{eq:second-effective-diffusion}) which is
\begin{equation}\label{eq:start}
\kappa^{ij}_{(2)}=-(n_0, (\tilde{\gamma}^{-1})^{i}_{l}F^{l}\chi^j),
\end{equation}
where $\chi^j$ is given in (\ref{GeneralAuxEq}).

Under the assumptions made the second term on the right hand side of this equation
($\partial_j(B^{kj}_{11}m_0)$) vanishes, and as discussed in Appendix~\ref{LocalChartappendix} 
the third term is proportional to $\alpha$ in the local coordinates. By making 
the local patches sufficiently small (evaluating the right hand side sufficiently close 
to the reference point in the center of patch) this term can therefore also be neglected,
and the auxiliary equation to solve is more simply
\begin{equation}
\label{Aux1}
\mathcal{L}_0^{\dagger}\chi^{i} =m_0\left( (\gamma^{-1})^{i}_lF^{l} - V^{i}\right).
\end{equation}
As $\vec{F}$ does not depend on the small scales we can look for solutions through a secondary 
auxiliary equation
\begin{equation}
{\cal L}_0^{\dagger} \lambda^{ij} = (\tilde{\gamma}^{-1})^{ij}.
\label{eq:A}
\end{equation}
where $\lambda$ is a tensor and $\tilde{\gamma}^{-1}$ is the traceless part of the mobility matrix. 
We note that this quantity is an element of the irreducible representation $\underline{\mathbf{5}}$
of $SO(3)$ which transforms as
\begin{equation}
\tilde{\gamma}^{-1} = R^*r(\alpha)\tilde{\gamma}^{-1}_B  r^{-1}(\alpha)(R^*)^{-1},
\label{eq:gamma-B}
\end{equation} 
where $\tilde{\gamma}^{-1}_B$ is the traceless part of the mobility tensor in the body frame.

We now consider the operator $\mathcal{L}_0^{\dagger}$ in (\ref{OperatorMdagger}) acting on $\lambda$.
As we have assumed no dependence on the small spatial scales, no cross-diffusion and no torque, the only term that matters
are the partial derivatives with respect to the rotations.
For a small patch the terms $\frac{1}{2}\sqrt{g}b_{22}^{ab}$ will be close to $k_B T (\eta^{-1})^{ab}$
and as discussed in Appendix~\ref{LocalChartappendix} we can interchange the order of the derivative and $\frac{1}{2}\sqrt{g}b_{22}^{ab}$
such that 
\begin{equation}
\mathcal{L}_0^{\dagger} = k_B T (\eta^{-1})^{ab}\frac{\partial^2}{\partial\alpha^{a} \partial\alpha^{b}} .
\label{eq:A-2}
\end{equation}
We then make the ansatz that $\lambda$ also lies in $\underline{\mathbf{5}}$:
\begin{equation}
\lambda = R^*r(\alpha)Q_B r^{-1}(\alpha)(R^*)^{-1},
\label{eq:ansatz}
\end{equation}
where $Q_B$ is an auxiliary traceless symmetric tensor in the body frame.
Note that if $Q^{s}_{B}$ would have a component $q\mathbf{1}$ where $q$ is a scalar,
then ${\cal L}^{\dagger}Q^{s}_{B}$ would be zero, and furthermore it would not contribute to (\ref{eq:start}).
We therefore prefer to solve (\ref{eq:A}) for a general symmetric matrix $Q_B$, then show that these solutions are only determined
up to an arbitrary term proportional to the identity matrix, after which by adding 
or subtracting such a term we can always adjust the trace of $Q_B$ to be zero.

The equation to solve is thus
\begin{widetext}
\begin{equation}
 \left[k_BT(\eta^{-1})^{mn}\frac{\partial}{\partial \alpha^m}\frac{\partial}{\partial \alpha^n} \left( R^*r(\alpha)Q_B r^{-1}(\alpha)(R^*)^{-1} \right)^{ij}\right]_{\alpha=0} =  (\tilde{\gamma}^{-1})^{ij}. 
\label{eq:ansatz-v2}
\end{equation} 
\end{widetext}
The pertinent terms inside the differentials are the quadratic, and using (\ref{eq:R}) one obtains:
\begin{widetext}
\begin{equation}\label{eq:ansatz-v3}
\begin{split}
&(\tilde{\gamma}^{-1})^{ij} = k_B T\bigg( (3R^*Q_B\eta^{-1}(R^*)^{-1} + 3R^*\eta^{-1}Q_B(R^*)^{-1}- 4\hbox{Tr}(\eta^{-1})R^*Q_B(R^*)^{-1}- 2\hbox{Tr}(Q_B)R^*\eta^{-1}(R^*)^{-1})^{ij}+\\
&+2(\hbox{Tr}(\eta^{-1})\hbox{Tr}(Q_B) -\hbox{Tr}(\eta^{-1}Q_B))\mathbf{1}^{ij} \bigg).
\end{split}
\end{equation}
\end{widetext}
It is seen that the factors $R^*$ and $(R^*)^{-1}$ can be removed on both sides.

The \textit{diagonal} elements obey coupled equations
\begin{widetext}
\begin{equation}
\left(\begin{array}{c} \frac{\tilde{\gamma}^{-1}_{B_{11}}}{2k_BT} \\  \frac{\tilde{\gamma}^{-1}_{B_{22}}}{2k_BT} \\  \frac{\tilde{\gamma}^{-1}_{B_{33}}}{2k_BT} \end{array}\right)
=\left(\begin{array}{ccc}
                   2(\eta^{-1}_{11}-a) & -\eta^{-1}_{22}     & -\eta^{-1}_{33}\\
                   -\eta^{-1}_{11}     & 2(\eta^{-1}_{22}-a) & -\eta^{-1}_{33} \\
                   -\eta^{-1}_{11}     & -\eta^{-1}_{22}     & 2(\eta^{-1}_{33}-a) \end{array}\right)
                   \left(\begin{array}{c} Q_{B_{11}}\\ Q_{B_{22}}\\ Q_{B_{33}}\ \end{array}\right).             
\label{eq:ansatz-v3-diagonal}
\end{equation}
\end{widetext}
The general solution of
(\ref{eq:ansatz-v3-diagonal})
can be written as
\begin{equation}
Q_{B_{ii}}=-\frac{1}{6b k_BT}\left(3\eta^{-1}_{ii}\tilde{\gamma}^{-1}_{B_{ii}} - \sum_j\eta^{-1}_{jj}\tilde{\gamma}^{-1}_{B_{jj}}\right) + q,
\label{eq:ansatz-v4-diagonal}
\end{equation}
where $b=(\eta^{-1}_{11}\eta^{-1}_{22}+\eta^{-1}_{22}\eta^{-1}_{33}+\eta^{-1}_{11}\eta^{-1}_{33})$ and $q$ is arbitrary.
By choosing $q=0$ we adjust the trace of $Q_B$ to be zero.
When $\tilde{\gamma}^{-1}_{B}$ commutes with $\eta^{-1}$ (\ref{eq:ansatz-v4-diagonal}) is the full result,
but in the more general case $\tilde{\gamma}^{-1}_{B}$ is not diagonal in the chosen basis and the off-diagonal elements of $Q_B$ are:
\begin{equation}
\label{eq:ansatz-v4-off-diagonal}
 Q_{B_{ij}}=-\frac{\tilde{\gamma}^{-1}_{B_{ij}}}{(k_BT)(4\hbox{Tr}(\eta^{-1})-3\eta^{-1}_{ii} - 3\eta^{-1}_{jj})} \,\, i\neq j 
\end{equation}
Using the invariant method, we obtain the general formula for the effective diffusion tensor:
\begin{widetext}
\begin{equation}\label{Kcomplete}
(K^{eff})^{ik}=\kappa^{ik}_{(1)}+\kappa^{ik}_{(2)}=\delta^{ik}\frac{k_B T \hbox{Tr}(\gamma^{-1})}{3} + \hbox{Tr}(\tilde{\gamma}^{-1}_{B}Q_B)\sum_{l,j}\bigg[\frac{1}{30}(\delta^{ij} \delta^{kl}F_{l}F_{j})+\frac{1}{10}(\delta^{ik}\delta^{lj}F_{l}F_{j}) \bigg] \, ,
\end{equation}
where
\begin{equation}
\label{eq:TrgammaQ}
 \hbox{Tr}(\tilde{\gamma}^{-1}_{B}Q_B)=\frac{\sum_{i}(\tilde{\gamma}^{-1}_{B})^{ii}(\tilde{\gamma}^{-1}_{B})^{ii}(\eta^{-1})^{ii}}{2bk_BT}+ \frac{2}{k_{B}T}\bigg(\frac{((\tilde{\gamma}^{-1}_{B})_{21})^2}{\hbox{Tr}(\eta^{-1})+3\eta^{-1}_{33}} + \frac{((\tilde{\gamma}^{-1}_{B})_{23})^2}{\hbox{Tr}(\eta^{-1})+3\eta^{-1}_{11}} + \frac{((\tilde{\gamma}^{-1}_{B})_{13})^2}{\hbox{Tr}(\eta^{-1})+3\eta^{-1}_{22}} \bigg) 
\end{equation}
\end{widetext}
In the special case where $\tilde{\gamma}^{-1}_{B}$ commutes with $\eta^{-1}$ only the first term in
(\ref{eq:TrgammaQ}) is non-zero which can be seen to be the result derived in 
\cite{Brennerpaper} and \cite{5Authors}.

\subsection{General potential motion on $SO(3)$}
\label{s:general-potential}
We now return to the setting of (\ref{s:semi-analytic-solution}) and assume that there is another
set of torques which satisfy a detailed balance condition but with a non-trivial zeroth probability $P^{(0)}=\sqrt{g}e^{-V}/\mathcal{N}_{(V)}$.
Using (\ref{eq:second-effective-diffusion-V}) the contribution to the effective diffusivity is
\begin{equation}
\label{eq:general-potential-formula}
\kappa^{ik}_{(2)}(V)= {T^{ik}}_{jl} F^j F^l
\end{equation}
where the fourth order tensor is given by
\begin{equation}
\label{eq:general-potential-formula}
{T^{ik}}_{jl} = - \int\sqrt{g} (\tilde{\gamma}^{-1})^i_j (\mathcal{L}_V^{\dagger})^{-1}\left[ (\tilde{\gamma}^{-1})^k_l\frac{1}{\mathcal{N}_{(V)}}e^{-V}\right]
\end{equation}
%with \textit{the same} tensor $\lambda$ given by
%(\ref{eq:ansatz}), (\ref{eq:ansatz-v4-diagonal}) and (\ref{eq:ansatz-v4-off-diagonal}) above.
with a tensor different from the equations (\ref{eq:ansatz}), (\ref{eq:ansatz-v4-diagonal}) and (\ref{eq:ansatz-v4-off-diagonal}) above. 
%We will comment on this semi-analytical result in Discussion.

\section{Motion on $\mathbb{C}^2$ and transformations in the Lie group $SU(2)$}
\label{s:su2}
As a further example we consider motion in four real dimensions with an internal symmetry different from the four-dimensional rotation group $SO(4)$.
Let four real coordinates $(x_1,y_1,x_2,y_2)$ be grouped into two complex coordinates
 $(z_1,z_2)$. 
An overdamped dynamics in  $(x_1,y_1,x_2,y_2)$, analogous to the first line of (\ref{Overdampedequations-physical-diagonal}),
can then be taken to be
\begin{equation}
\label{eq:2C-dynamics}
\begin{split}
&\left(\begin{array}{c}
                     \text{d}z_1   \\
                     \text{d}z_2 \end{array}\right)\,=\,\mathbf{S}	\left(\begin{array}{c}
                     f_1   \\
                     f_2 \end{array}\right)\,\text{d}t+\sqrt{2k_BT}\,\mathbf{S}^{\frac{1}{2}}\,\bullet \left(\begin{array}{c}
                     \text{d}\omega_1   \\
                     \text{d}\omega_2 \end{array}\right)\\
% &\text{d}\vec{\alpha}=\,\mathbf{\eta}^{-1} \vec{M}\text{d}t+\sqrt{2k_BT}\, \mathbf{\eta}^{-\frac{1}{2}}\bullet\text{d}\vec{\xi}
 %\begin{widetext}
%\end{widetext}
\end{split}
\end{equation}
where $\mathbf{S}$ is a Hermitian mobility matrix,
$f_1$ and $f_2$ are two complex forces and $\text{d}\omega_1$, and $\text{d}\omega_2$ are two complex Wiener process such that
\begin{equation}
\begin{split}
&\left<\text{d}\omega_{\mu}(t)\text{d}\omega_{\nu}(s) \right>=0;\\
&\left<\text{d}\omega_{\mu}(t)\overline{\text{d}\omega_{\nu}}(s)\right>=2\text{d}t\delta_{{\mu}{\nu}}\delta(t-s);
\end{split}
\end{equation}
Two-dimensional complex space has the symmetry of the \textit{special unitary group} ($SU(2)$) of $2 \times 2$ unitary matrices with determinant $1$.
To have diffusion on $\mathbb{C}^2$ with the internal symmetry of $SU(2)$ we should therefore add to (\ref{eq:2C-dynamics}) 
\begin{equation}
\begin{split}
\label{eq:SU2-dynamics}
 &\text{d}\vec{\alpha}=\,\mathbf{\eta}^{-1} \vec{M}\text{d}t+\sqrt{2k_BT}\, \mathbf{\eta}^{-\frac{1}{2}}\bullet\text{d}\vec{\xi}
 %\begin{widetext}
%\end{widetext}
\end{split}
\end{equation}
where $\alpha$ is the coordinate in some local chart of $SU(2)$
and the mobility tensor transforms as 
\begin{equation}
\label{eq:SU2-transformation}
\mathbf{S}(g)=U(g)\mathbf{S}(e)U^{-1}(g)
\end{equation} 
Eqs.~(\ref{eq:2C-dynamics}-\ref{eq:SU2-transformation}) define the model we will study in this section.

The following facts about $SU(2)$ are well known 
\begin{itemize}
\item If  $z=(z_1,z_2)$ and  $w=(w_1,w_2)$ are two pairs of complex numbers $SU(2)$ preserves the scalar product 
$(z,w)=\overline{z_1}w_1+\overline{z_2}w_2$. Introducing the covariant vector (``ket'') $z^i$ and the contravariant vector (``bra'')
$z_i=\overline{z^i}$  $SU(2)$ hence preserves the identity operator $\mathbf{1}^i_j$.
In particular  $SU(2)$ preserves the norm $|z|^2=(z,z)=\sum_i z_i z^i$.
\item $SU(2)$ is a compact manifold of dimension $3$, and is the two-fold covering group of $SO(3)$. 
\item The tangent space of $SU(2)$ at the identity element $e$ is the Lie algebra $su(2)$, isomorphic to $\mathbb{R}^3$ (and to $so(3)$).
An element of the Lie algebra can be written $v=i\sum_i \alpha_i \sigma_i$ where $(\alpha_1,\alpha_2,\alpha_3)$ are three real parameters
and $(\sigma_1,\sigma_2,\sigma_3)$ are three Hermitian traceless matrices, commonly chosen to be the Pauli matrices
\begin{equation}\label{paulimatrices}
\sigma_1=	\left(\begin{array}{cc}
                   0 & 1    \\
                   1 & 0 \end{array}\right)					\,\,,\, \sigma_2=	\left(\begin{array}{cc}
                   0 & -i    \\
                   i & 0 \end{array}\right)				\,\,,\,\sigma_3=\left(\begin{array}{cc}
                   1 & 0    \\
                   0 & -1 \end{array}\right)			\,.
\nonumber
\end{equation}  
\item The exponential map of the Lie algebra covers the group. If $H$ is a Hermitian $2 \times 2$ matrix with zero trace then $U=e^{iH}$ is in $SU(2)$.  
\end{itemize} 

In this setting the analysis proceeds much as above for $SO(3)$. Qualitatively speaking one main difference is that $SU(2)$ only preserves the identity matrix
$\mathbf{1}^i_j$, and not, as $SO(3)$, both an identity matrix and a three-dimensional Euclidean metric. A simplifying feature is on the other hand 
that the friction operator $\mathbf{S}$ can be expressed in the same basis as the Lie algebra.
Let us therefore write
\begin{equation}
\mathbf{S}=\mu_{0}\mathbf{I}+\sum_{l}\mu_{l}\sigma_{l}\qquad \text{Tr}(\mathbf{S})=2\mu_0.
\end{equation}
The advection velocity is then
\begin{equation}
	V^{\mu} = \mu_0 f^{\mu}
\end{equation}
We also introduce the trace-less mobility tensor
\begin{equation}
{\tilde{\mathbf{S}}^{\mu}}_{\nu}=\sum_{l}\mu_{l}{(\sigma_{l})^{\mu}}_{\nu}
\end{equation}
The analogue of the auxiliary tensor field $\lambda$ for $SO(3)$ above is now a Hermitian $2\times 2$ matrix,
and the analogue of the ansatz (\ref{eq:ansatz}) is
\begin{equation}
{\lambda^{\mu}}_{\nu} = {U^{\mu}}_{\lambda} {(Q_B)^{\lambda}}_{\kappa} {(U^{-1})^{\kappa}}_{\nu}
\end{equation}
As $Q_B$ is Hermitian it can be written
\begin{equation}
Q_B=q_{0}\mathbf{I}+\sum_{l}q_{l}\sigma_{l}
\end{equation}
and the analogue of (\ref{eq:ansatz-v2}) is
\begin{widetext}
\begin{equation}\label{equationauxiliarySU(2)}
\begin{split}
k_BT\sum_{i}(\eta^{-1})^{ii}\frac{\partial^{2}}{\partial \alpha^{2}_i}\left\{U(\vec{\alpha})\left(q_0\mathbf{1} +\sum_{k} q_{k}\sigma_{k}\right) U^{-1}(\vec{\alpha}) \right\}=-U(\vec{\alpha})\sum_{k}\mu_k\sigma_k U^{-1}(\vec{\alpha}),
\end{split}
\end{equation}
\end{widetext}
with solution
\begin{equation}\label{solutioninSU(2)}
q_{p}=\frac{1}{4k_BT}\frac{\mu_p}{\text{Tr}(\eta^{-1})-\eta^{-1}_{pp}}
\end{equation}
and $q_{0}$ arbitrary.

The effective diffusion on $C^2$ is determined by the fourth-order tensor
\begin{equation}
{T^{\mu\nu}}_{\kappa\lambda}=\left< {\tilde{\mathbf{S}}^{\mu}}_{\kappa} {\lambda^{\nu}}_{\lambda} \right>
\end{equation}
which by $SU(2)$ invariance can be written $A \mathbf{1}^{\mu}_{\kappa}\mathbf{1}^{\nu}_{\lambda} + B \mathbf{1}^{\mu}_{\lambda}\mathbf{1}^{\nu}_{\kappa}$.
The coefficients are determined by the two equations
\begin{eqnarray}
0&=&4A+2B \\
\text{Tr}\left[\tilde{\mathbf{S}}\lambda\right]&=& 2A+4B
\end{eqnarray}
and it therefore suffices to compute
\begin{equation}
\mathcal{B}=4k_B T \text{Tr}\left[\tilde{\mathbf{S}}\lambda\right] = \sum_{p}\frac{\mu^{2}_{p}}{\text{Tr}(\eta^{-1})-(\eta^{-1})^{pp}}.
\end{equation}
The effective diffusion has two components of which the first is
\begin{widetext}
\begin{equation}
2 \left(\text{Re}(\nabla^{2}_{z_\mu \overline{z}_\mu})\right)\left(\kappa^{\mu \mu}_{(2)}C^{(0)}\right)=\left(\nabla_{x_\mu}\nabla_{x_\mu}+\nabla_{y_\mu}\nabla_{y_\mu}\right)\left(C^{(0)}\frac{\mathcal{B}}{3k_BT}f^\mu f^\mu \right).
\end{equation}
\end{widetext}
and the second is
\begin{widetext}
\begin{equation}
\text{Re}\left(\nabla^{2}_{z_\mu z_\nu}\kappa^{\mu \nu}_{(2)}C^{(0)}-\nabla^{2}_{\overline{z}_\mu \overline{z}_\nu}\kappa^{\mu \overline{\nu}}_{(2)}C^{(0)}\right)=\text{Re}\left(\nabla^{2}_{z_\mu z_\nu}\frac{\mathcal{B}}{6k_BT}f^\mu f^\nu C^{(0)}-\nabla^{2}_{\overline{z}_\mu \overline{z}_\nu}\frac{\mathcal{B}}{6k_BT}f^\mu \overline{f}^\nu C^{(0)}\right).
\end{equation}
\end{widetext}

\section{Numerical Simulations}
\label{s:secsim}

In this section we compare our result (\ref{Kcomplete}) to numerical simulations of a Brownian particle, which can translate and rotate in $3$-dimensions and  subjected to an external constant force field.
During the simulation we solve numerically the equations in (\ref{Overdampedequations-physical-diagonal}), where the evolution of translational Brownian motion is computed in the lab frame, while the rotational Brownian motion is computed in the body frame. We also assume that:
\begin{itemize}
\item  The translational friction tensor is only dependent by orientation in the lab frame.
\item  The rotational friction tensor and the translational friction tensor are not diagonal in the same base in the body frame.
\item  The force is constant in the lab frame and is applied only in one direction.
\item  The torque is set to zero.
\end{itemize}
  
For the simulations, we choose quaternion as a concrete parameterisation of rotation, i.e. the model is supplemented by the equation of motion for the quaternion $q$,
\begin{equation}
\dot{q}=\frac{1}{2}q \circ \Omega \,,
\end{equation}
where the symbol $\circ$ denotes a \textit{quaternion} product \cite{quaternion} evaluated in the Stratonovich sense, and $\Omega$ is the angular velocity in the body frame (from the equation describing the rotational motion), represented as a pure quaternion \cite{quaternion}.

We solve translational and rotational equations describing the motion of Brownian body with the Euler algorithm with time-step $10^{-4}$ s and temperature $T=300\,\, K$.  
The result of the simulations averaged over $10^4$ realisations of the noise sources (with identical initial conditions) are shown in the following figures.

In Fig. \ref{figtraj1} a trajectory in lab frame of a Brownian body is shown. The shape of the particle is identified by translational friction tensor and rotational friction tensor in the body frame:
\begin{widetext}
\begin{equation}\label{tensorrottran}
\gamma_B=\left(\begin{array}{ccc}
25. & 0. & 0.2 \\
 0. & 16. & 0. \\
 0.2 & 0. & 13. \end{array}\right)\,\, [\frac{\text{fN}\,\text{s}}{\mu\text{m}}] \,\,\,\,\,\,, \eta=\left(\begin{array}{ccc}
 17. & 0. & 0. \\
 0. & 16. & 0. \\
 0. & 0. & 8. \end{array}\right)\,\, [\text{fN}\,\mu\text{m}\,\text{s}]. 
 \end{equation}
\end{widetext}

As expected the motion of the body is enhanced along the direction where the force is applied, while diffusion governs the motion along the other two directions. In this specific case the force is applied along the \textit{x}-direction with magnitude equal to $\| \vec{F} \|=10.\,\text{fN}$. 

\begin{figure}
\centering
\includegraphics[width=1\columnwidth, keepaspectratio=true, angle=0]{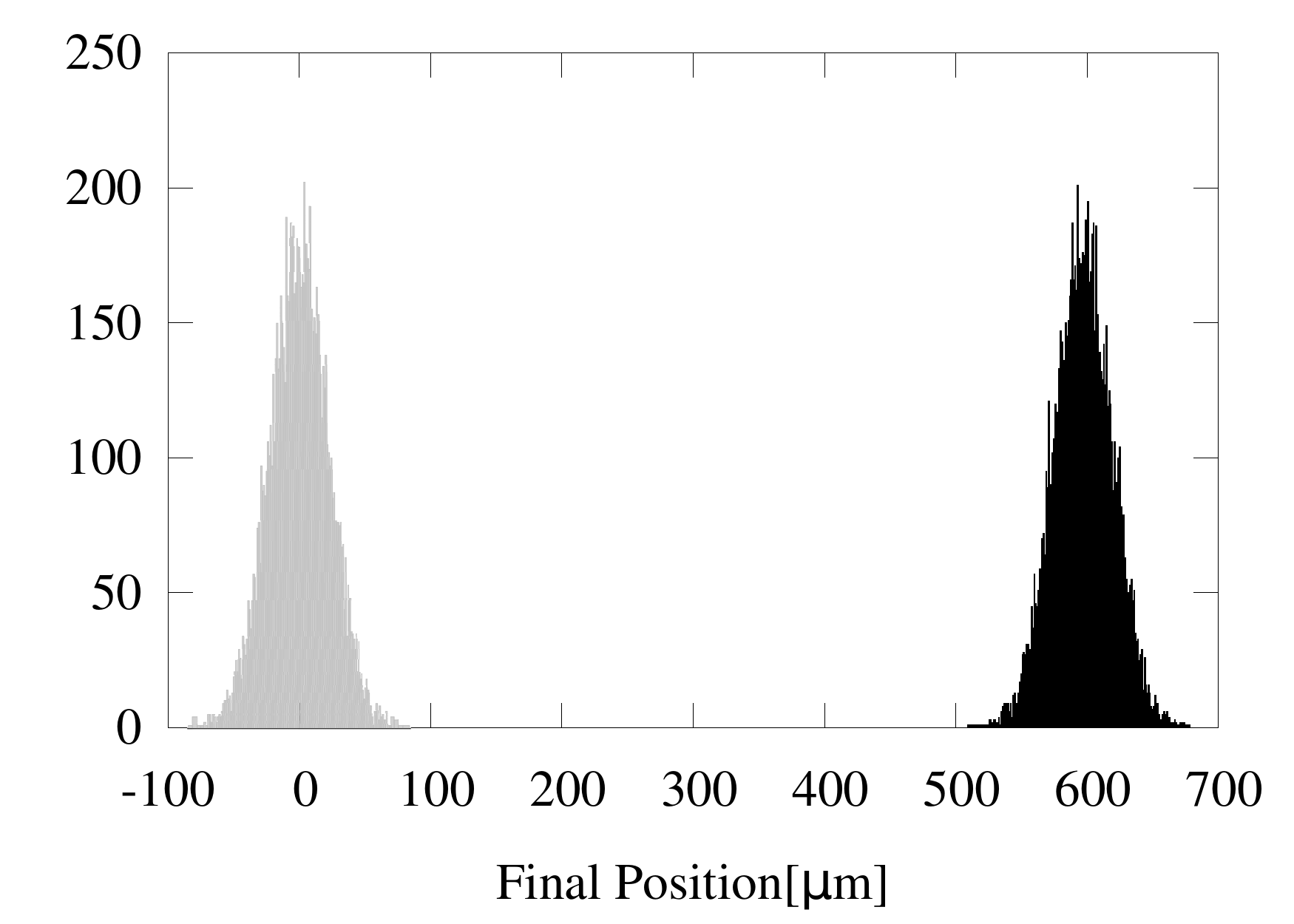}
\caption{Color online. The figure describes the final position distribution probability functions for the $x$- and $z$-component  in lab frame of all trajectories for a Brownian body, respectively (black) right and (grey) left probability distribution (the $y$-component probability distribution function is not shown. It overlaps with the $z$-component probability distribution function, as expected). The asymmetry is given by the application of an external force field on the body. Body: $(\gamma_B)^{11}=25.\,\,\, \frac{\text{fN s}}{\mu \text{m}}\, , (\gamma_B)^{22}=16.\,\,\, \frac{\text{fN s}}{\mu \text{m}}\,,(\gamma_B)^{33}=13.\,\,\, \frac{\text{fN s}}{\mu \text{m}}$, $(\gamma_B)^{13}=0.2\,\,\, \frac{\text{fN s}}{\mu \text{m}}\, (\gamma_B)^{12}=(\gamma_B)^{23}=0.\,\,\, \frac{\text{fN s}}{\mu \text{m}} $, $(\eta)^{11}=17.\,\,\,\text{fN}\, \mu \text{m} \,\text{s}\, ,(\eta)^{22}=16.\,\,\,\text{fN}\, \mu \text{m} \,\text{s}\, , (\eta)^{33}=8.\,\,\,\text{fN}\, \mu \text{m} \,\text{s}$. The external force is $\vec{F}=(10.\, \text{fN}, 0\, \text{fN},0\, \text{fN})$.}
\label{Figisto}
\end{figure}

Fig. \ref{Figisto}, instead, shows the probability distribution functions of the final position of all trajectories, for the same body described before. Here the asymmetry given by the application of an external force is well displayed. Indeed the final position  mean of Brownian particle, along the coordinate where the force is applied, can be computed as:
\begin{equation}
\left< \vec{x}_{final}\right>=\vec{V}_bt_{final},
\end{equation}
where $\vec{V}_b$ is given by equation (\ref{ballisticvelocitySO(3)}) and $t_{final}$ is set to $1000$ s.
In this case the ballistic velocity of the Brownian body is $V^{x}_b=0.60\,\frac{\mu \text{m}}{\text{s}}$. The  final position mean is therefore $\left< x_{final}\right>=600.\,\, {\mu}\text{m}$.

\begin{figure}
\centering
\includegraphics[width=1\columnwidth, keepaspectratio=true, angle=0]{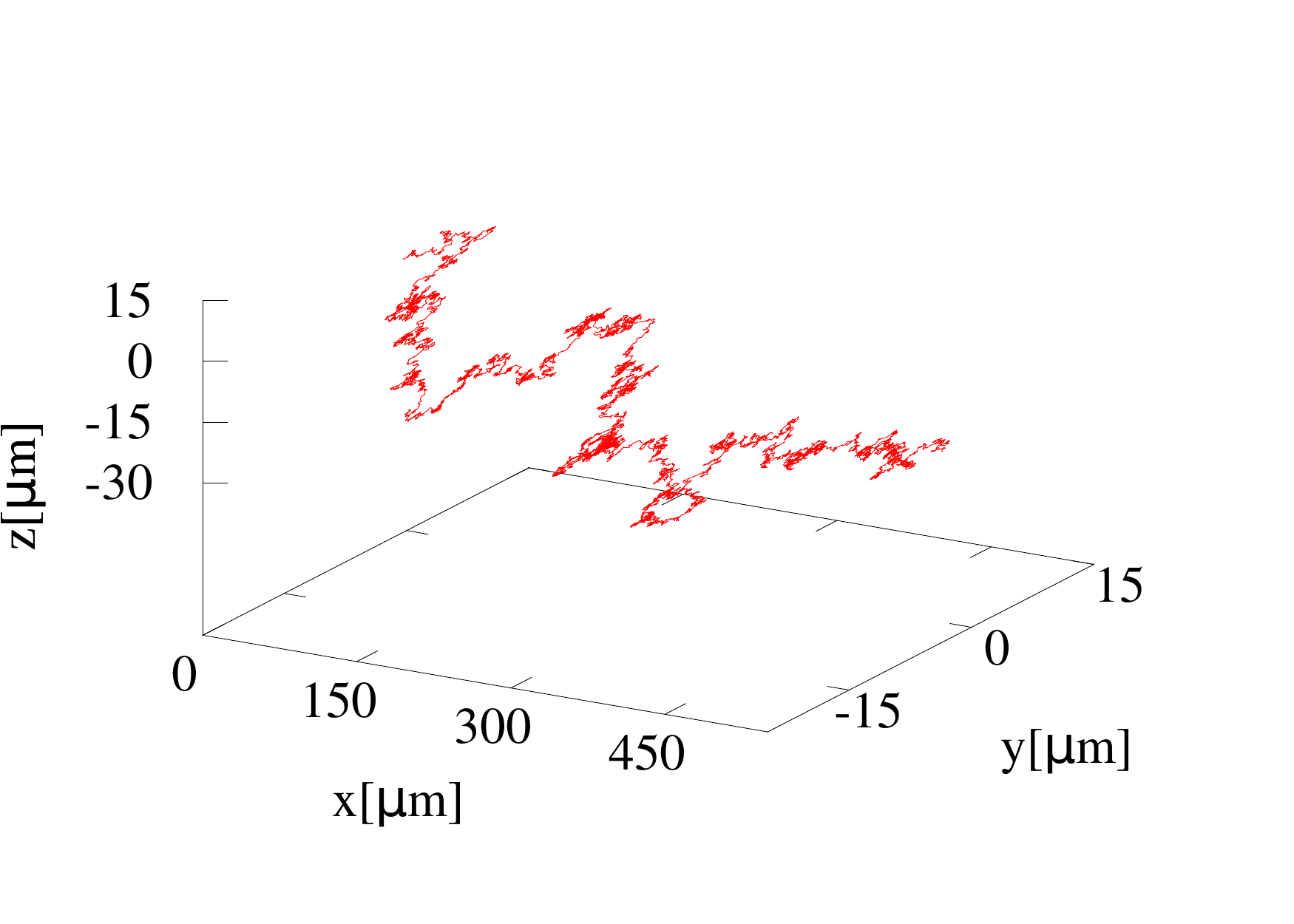}
\caption{Color online. The figure describes a trajectory in lab frame of a Brownian body in three dimensions. Body: $(\gamma_B)^{11}=25.\,\,\, \frac{\text{fN s}}{\mu \text{m}}\, , (\gamma_B)^{22}=16.\,\,\, \frac{\text{fN s}}{\mu \text{m}}\,,(\gamma_B)^{33}=13.\,\,\, \frac{\text{fN s}}{\mu \text{m}}$, $(\gamma_B)^{13}=0.2\,\,\, \frac{\text{fN s}}{\mu \text{m}}\, (\gamma_B)^{12}=(\gamma_B)^{23}=0.\,\,\, \frac{\text{fN s}}{\mu \text{m}} $, $(\eta)^{11}=17.\,\,\,\text{fN}\, \mu \text{m} \,\text{s}\, ,(\eta)^{22}=16.\,\,\,\text{fN}\, \mu \text{m} \,\text{s}\, , (\eta)^{33}=8.\,\,\,\text{fN}\, \mu \text{m} \,\text{s}$. The external force is $\vec{F}=(10.\, \text{fN}, 0\, \text{fN},0\, \text{fN})$.}
\label{figtraj1}
\end{figure}

\begin{figure}
\centering
\includegraphics[width=1\columnwidth, keepaspectratio=true, angle=0]{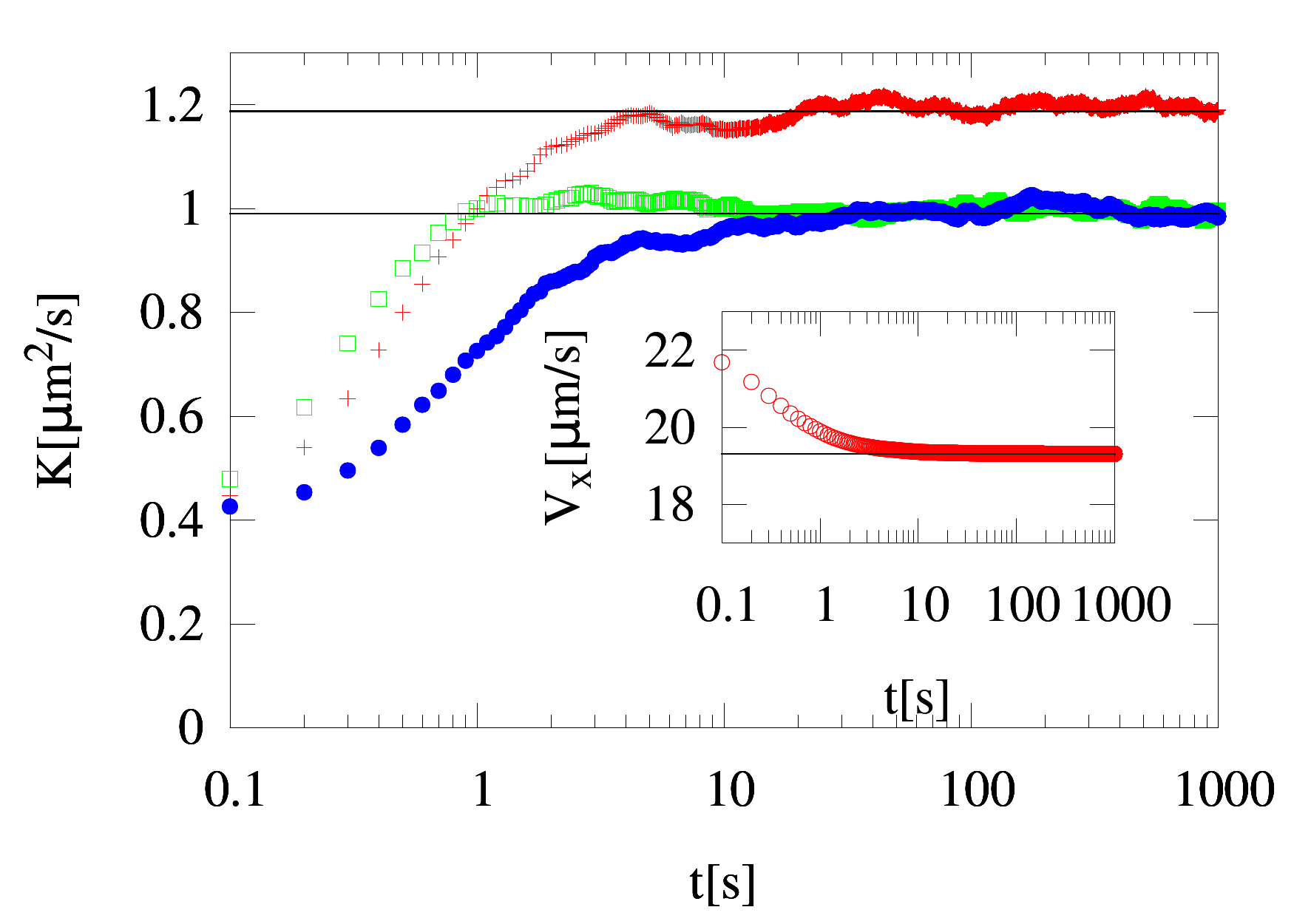}
\caption{Color online. The figure describes the diffusion coefficient along the \textit{x}-axis, \textit{y}-axis and \textit{z}-axis in lab frame, respectively (red) cross, (green) empty square, and (blue) full circle points, as function of the time 
for a body at room temperature $T=300\,\,\, K$, averaged over $10^4$ realisation of the Gaussian noise source, but for identical initial position and orientation of the body. 
The two full horizontal lines represent the theoretical long-term prediction from (\ref{Kcomplete}), $(K^{eff})^{xx}=1.18 \mu \text{m}^2/\text{s}$, $(K^{eff})^{yy}=(K^{eff})^{zz}=0.99 \mu \text{m}^2/\text{s}$. 
Body: $(\gamma_B)^{11}=12.\,\,\, \frac{\text{fN s}}{\mu \text{m}}\, , (\gamma_B)^{22}=(\gamma_B)^{33}=10.\,\,\, \frac{\text{fN s}}{\mu \text{m}}$, $(\gamma_B)^{12}=2.\,\,\, \frac{\text{fN s}}{\mu \text{m}}\, (\gamma_B)^{13}=(\gamma_B)^{23}=0.\,\,\, \frac{\text{fN s}}{\mu \text{m}} $, $(\eta)^{11}=7.12\,\,\,\text{fN}\, \mu \text{m} \,\text{s}\, ,(\eta)^{22}=6.84\,\,\,\text{fN}\, \mu \text{m} \,\text{s}\, , (\eta)^{33}=4.12\,\,\,\text{fN}\, \mu \text{m} \,\text{s}$. The external force is $\vec{F}=(200\, \text{fN}, 0\, \text{fN},0\, \text{fN})$. In the inset is plotted the $x$-component of the ballistic velocity of the particle (red) empty circle points, the full horizontal line is equal to $V^x=19.31 \mu \text{m}/\text{s}$ 
(the other net velocity components are zero,  not shown).}
\label{figeffdiff1}
\end{figure}

\begin{figure}
\centering
\includegraphics[width=1\columnwidth, keepaspectratio=true, angle=0]{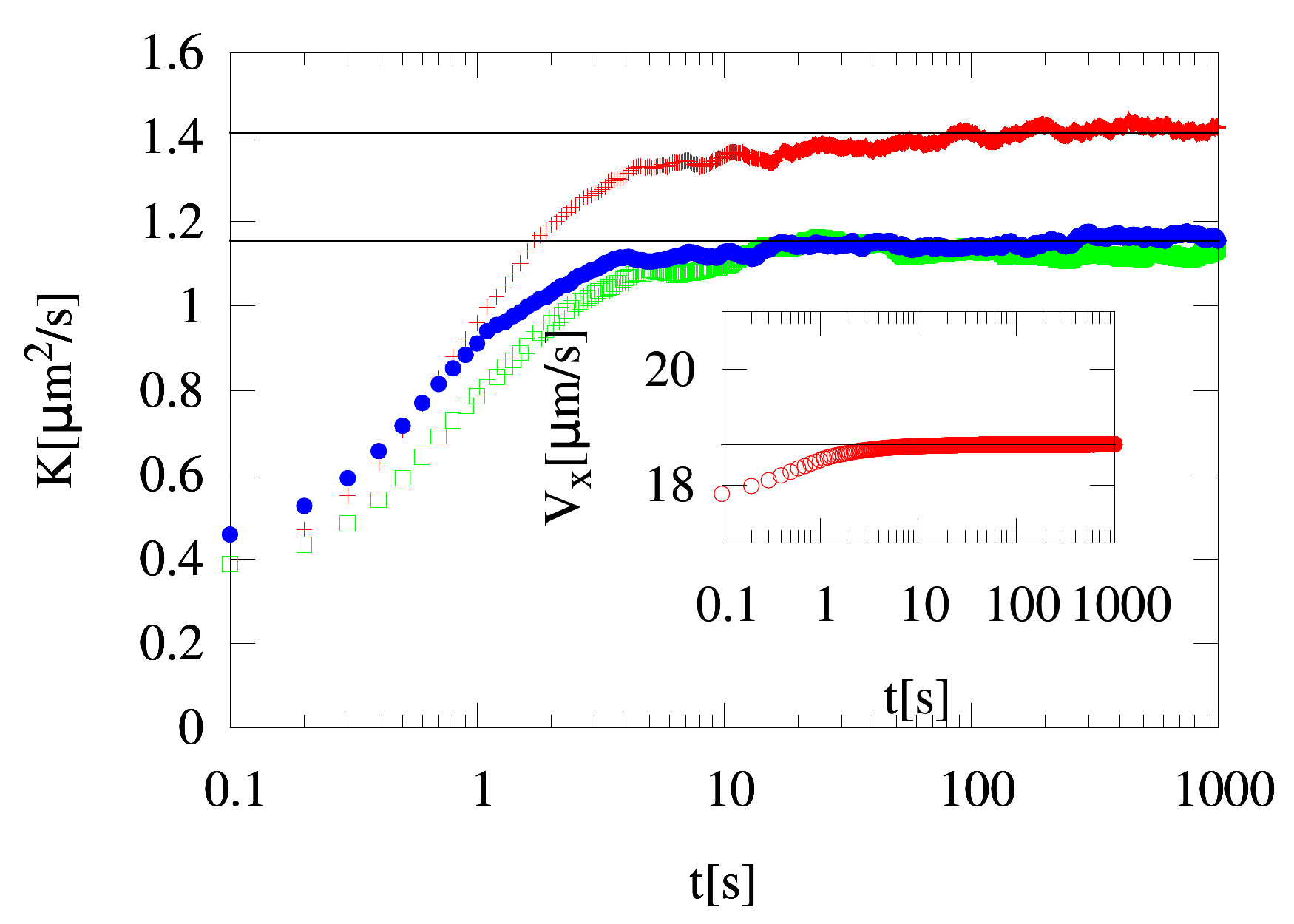}
\caption{Color online. The figure describes the diffusion coefficient along the \textit{x}-axis, \textit{y}-axis and \textit{z}-axis in lab frame, respectively (red) cross, (green) empty square, and (blue) full circle points, as function of the time 
for a body at room temperature $T=300\,\,\, K$, averaged over $10^4$ realisation of the Gaussian noise source, but for identical initial position and orientation of the body. The two full horizontal lines represent the theoretical long-term prediction from (\ref{Kcomplete}), $(K^{eff})^{xx}=1.41\mu \text{m}^2/\text{s}$, $(K^{eff})^{yy}=(K^{eff})^{zz}=1.15 \mu \text{m}^2/\text{s}$. 
Body: $(\gamma_B)^{11}=12.\,\,\, \frac{\text{fN s}}{\mu \text{m}}\, , (\gamma_B)^{22}=11.\,\,\, \frac{\text{fN s}}{\mu \text{m}}\, ,(\gamma_B)^{33}=10.\,\,\, \frac{\text{fN s}}{\mu \text{m}}$, $(\gamma_B)^{12}=2.\,\,\, \frac{\text{fN s}}{\mu \text{m}}\, (\gamma_B)^{13}=(\gamma_B)^{23}=0.\,\,\, \frac{\text{fN s}}{\mu \text{m}} $, $(\eta)^{11}=7.12\,\,\,\text{fN}\, \mu \text{m} \,\text{s}\, ,(\eta)^{22}=6.84\,\,\,\text{fN}\, \mu \text{m} \,\text{s}\, , (\eta)^{33}=4.12\,\,\,\text{fN}\, \mu \text{m} \,\text{s}$. The external force is $\vec{F}=(200\, \text{fN}, 0\, \text{fN},0\, \text{fN})$. In the inset is plotted the $x$-component of the ballistic velocity of the particle (red) empty circle points, the full horizontal line is equal to $V^x=18.7 \mu \text{m}/\text{s}$ (the other net velocity components are zero,  not shown).}
\label{figeffdiff2}
\end{figure}

In Fig.\ref{figeffdiff1} we show the temporal evolution of the translational diffusion tensor  along the three different axes and, in the inset, the ballistic velocity, component \textit{x}, for a Brownian body.  The components of the effective diffusion tensor computed in equation (\ref{Kcomplete}) are in very good agreement with the long-term coefficients obtained by the simulation. The external force applied to the system is $\vec{F}=(200\, \text{fN}, 0\, \text{fN},0\, \text{fN})$, while the translational friction tensor and rotational friction tensor in the body frame have the following form:
\begin{widetext}
\begin{equation}\label{tensorrottran}
\gamma_B=\left(\begin{array}{ccc}
12. & 2.0 & 0. \\
 2.0 & 10. & 0. \\
 0. & 0. & 10. \end{array}\right)\,\, [\frac{\text{fN}\,\text{s}}{\mu\text{m}}] \,\,\,\,\,\,, \eta=\left(\begin{array}{ccc}
 7.12 & 0. & 0. \\
 0. & 6.84 & 0. \\
 0. & 0. & 4.12 \end{array}\right)\,\, [\text{fN}\,\mu\text{m}\,\text{s}]. 
 \end{equation}
\end{widetext}
The two full horizontal lines represent the theoretical long term prediction given by equation (\ref{Kcomplete}):
\begin{equation}
\mathbf{K}^{eff}=\left(\begin{array}{ccc}
1.18 & 0. & 0. \\
 0. & 0.99 & 0. \\
 0. & 0. & 0.99 \end{array} \right)\,\,[\frac{\mu\text{m}^2}{\text{s}}]. 
\end{equation}
The inset shows the evolution of the $x$-component of the ballistic velocity of the body. The full horizontal line is the long term prediction computed by equation (\ref{ballisticvelocitySO(3)}) and it is equal to $V^{x}=19.31 \, \frac{\mu\text{m}}{s}$.

Fig.\ref{figeffdiff2} shows the temporal evolution of the translational diffusion tensor  along the three different axes and, in the inset, the component \textit{x} of the ballistic velocity of a Brownian body.  The components of the effective diffusion tensor computed in equation (\ref{Kcomplete}) are in very agreement with the long-term coefficients obtained by the simulation. The external force applied to the system is $\vec{F}=(200\, \text{fN}, 0\, \text{fN},0\, \text{fN})$ and the translational friction tensor in the body frame has the following form:
\begin{equation}
\gamma_B=\left(\begin{array}{ccc}
12. & 2.0 & 0. \\
 2.0 & 11. & 0. \\
 0. & 0. & 10. \end{array}\right)\,\, [\frac{\text{fN}\,\text{s}}{\mu\text{m}}] \,\,\,\,\,\,.
 \end{equation}
The rotational friction tensor is the same of equation (\ref{tensorrottran}).
The two full horizontal lines represent the theoretical long term prediction given by equation (\ref{Kcomplete}):
\begin{equation}
\mathbf{K}^{eff}=\left(\begin{array}{ccc}
1.41 & 0. & 0. \\
 0. & 1.15 & 0. \\
 0. & 0. & 1.15 \end{array} \right)\,\,[\frac{\mu\text{m}^2}{\text{s}}]. 
\end{equation}
The inset shows the evolution of the $x$-component of the ballistic velocity of the body. The full horizontal line is the long term prediction computed by equation (\ref{ballisticvelocitySO(3)}) and it is equal to $V^{x}=18.70 \, \frac{\mu\text{m}}{s}$.

\section{Discussion}
\label{s:discussion}
In this paper we have analysed the coupled overdamped motion in space and on a manifold of internal states.
We have shown that on large scales in space and time the probability distribution over the internal states
and on small scales in space is slaved to the large-scale spatial probability  distribution, which is that of an
advection-diffusion process. This process is described by an effective drift field and an effective
diffusion matrix, as is to be expected on physical grounds. The method to arrive at these results 
is the multi-scale formalism, which here amounts to solving auxiliary 
elliptic partial differential equations on the manifold, and computing weighted averages of these solutions, over the manifold.
In general the solutions of the auxiliary equations and the averages can only be found numerically.
For the drift field we identified settings including cross-mobility between space and internal states where particles 
that are acted upon by different internal drifts (torques), but are otherwise equivalent, may migrate at different speeds in space.

We then showed that when the drift field obeys a detailed balance condition with respect to the diffusion operator
the auxiliary equations must be solved to compute the effective diffusion.
 %are a Poisson equation on the manifold, all essentially equivalent.
%It therefore suffices to solve one of these auxiliary equations to solve all of them.
A condition for such a solution to exist is that the average drift velocity vanishes, 
and the analysis therefore has to carried out in a co-moving frame in space. We have assumed that this co-moving frame moves with constant speed;
effects of the speed of co-moving frames varying on large scales of space and time have not been considered in this work.

We then applied our general approach to the Brownian translations and rotations which is advection-diffusion on $R^3$ and on 
Lie group $SO(3)$ of rotations in three dimensions. We considered the special case of constant angular friction and no torque and showed that
the auxiliary equations then can be solved in closed form. The solution can be described as follows: the traceless
mobility tensor $\tilde{\gamma}^{-1}$ is a traceless symmetric $3\times 3$ matrix, hence an element of $SO(3)$ irrep $\underline{\mathbf{5}}$. 
Its value at a given group element $g$ is determined by its
value at the group identity $\tilde{\gamma}_B^{-1}$, which is the traceless part of the mobility tensor in the body frame, and a rotation matrix $R(g)$,
through the formula $\tilde{\gamma}^{-1}=R(g) \tilde{\gamma}_B^{-1} R^T(g)$.
We then find another element of irrep $\underline{\mathbf{5}}$, $\lambda=R(g) Q_B R^T(g)$ where $Q_B$ is a linear transform of $\tilde{\gamma}_B^{-1}$.
The effective diffusion matrix is then computed as an average of the product $\tilde{\gamma}^{-1} \lambda$ over a uniform measure on rotations (Haar measure).
As discussed in the main text these averages can be computed in closed form using invariant theory.
We compare our predictions to numerical simulations with excellent results. 

Compared to results reported recently~\cite{5Authors} we have here not assumed that
the spatial mobility matrix $\gamma_B$ commutes with the angular diffusion matrix in the body frame.
Hence we have here enlarged the set of cases where the effective diffusion can be computed analytically in closed form.
We have also shown that the effective diffusion of  Brownian translations and rotations, which
has the same angular diffusion and which is in detailed balance, can be computed when an analytic expression of the solution of eq. (\ref{eq:solution-first-order-2})  can be found. 
%More interestingly, we have shown that the effective diffusion of \textit{any} Brownian translation and rotations which
%has the same angular diffusion and which is in detailed balance can be computed from averages of 
%\textit{the same} product $A_1^k\chi_{(V)}^i$ discussed above.
When the measure over rotations is not uniform these averages cannot be computed using invariant theory
and would require further study. We hope to be able to return to this issue and other applications of the
methods developed here in future contributions.

\section*{Acknowledgements} 
We thank Stefano Bo and Marcelo Dias and Ralf Eichhorn
for many discussions and a stimulating collaboration leading up to this work, Aleksandr Zheltukhin for discussions about Lie groups and
Paolo Muratore-Ginanneschi for constructive comments on an earlier version of the manuscript.
This work was supported by the Swedish Science Council through grant 621-2012-2982.

\appendix

\section{Transformation properties of the H\"anggi-Klimontovich drift}
We want to show that 
\begin{equation}
D^{a} = A^{a}-\frac{1}{2\sqrt{g}}\partial_{\alpha^b}\left(\sqrt{g}b^{ab}\right)
\end{equation}
defined in Eq.~(\ref{eq:HK}) above transforms as a vector. Consider therefore
in a transformed coordinate system
\begin{equation}
\label{eq:transformed-1}
{D'}^{a} = {A'}^{a}-\frac{1}{2\sqrt{g'}}\partial_{{\alpha'}^b}\left(\sqrt{g'}{b'}^{ab}\right).
\end{equation}
Using the transformation properties of $A$, $b$, $\sqrt{g}$ and derivatives this is
\begin{eqnarray}
\label{eq:transformed-2}
{D'}^{a} &=& J^{a}_b {A}^{b}+\frac{1}{2}\frac{\partial^2{\alpha'}^a}{\partial\alpha^a\partial\alpha^b} b^{pq}\nonumber \\
        && - \frac{d}{2\sqrt{g}}(J^{-1})^q_b \frac{\partial}{\partial\alpha^q}\left(\sqrt{g}\frac{1}{d}J^a_pJ^b_r b^{pr}\right)
\end{eqnarray}
where $J^a_b =\partial{\alpha'}^a/\partial\alpha^b$ and $d=\det J$. Identifying terms this is
\begin{eqnarray}
\label{eq:transformed-3}
{D'}^{a} &=& J^{a}_b {D}^{b}+\frac{1}{2}\frac{\partial^2{\alpha'}^a}{\partial\alpha^a\partial\alpha^b} b^{pq}\nonumber \\
        && - \frac{d}{2}(J^{-1})^q_b b^{pr} \frac{\partial}{\partial\alpha^q}\left(\frac{1}{d}J^a_pJ^b_r \right)
\end{eqnarray}
The second line in above gives three terms
\begin{eqnarray}
\hbox{$2^{\hbox{nd}}$ line} &=& \frac{d}{2}(J^{-1})^q_b b^{pr} \frac{\partial d}{\partial\alpha^q} \frac{1}{d^2}J^a_pJ^b_r \nonumber \\
                         && - \frac{d}{2}(J^{-1})^q_b b^{pr} \frac{\partial^2 {\alpha'}^a}{\partial\alpha^q\partial\alpha^p}\frac{1}{d}J^b_r  \nonumber \\
                         && - \frac{d}{2}(J^{-1})^q_b b^{pr} \frac{\partial^2 {\alpha'}^b}{\partial\alpha^q\partial\alpha^r}\frac{1}{d}J^a_p 
\end{eqnarray}
Since $(J^{-1})^q_b J^b_r=\mathbf{1}^q_r$ the second line in above 
cancels with the second term in the first line of (\ref{eq:transformed-3}).
The first line in above can similarly be 
be re-written  $\frac{1}{2}  J^a_p b^{pq} \partial_{\alpha^q} \log d$. Since  $\partial_{\alpha^q} \log d=\hbox{Tr}\left[(J^{-1}\frac{\partial J}{\partial\alpha^q}\right]$
this cancels with the third line and we have hence simply
\begin{eqnarray}
\label{eq:transformed-4}
{D'}^{a} &=& J^{a}_b {D}^{b} + \hbox{all other terms cancel}
\end{eqnarray}

\section{Local charts of the $SO(3)$ group manifold} 
\label{LocalChartappendix}
Let $e$ be the unit element of $SO(3)$ and let $\mathbf{1}$ be its representation as a unit rotation matrix.
The matrices
\begin{equation}
\textbf{M}(\alpha)=\left(\begin{matrix}0 & -\alpha_3 & \alpha_2 \\ \alpha_3 & 0 & -\alpha_1 \\ -\alpha_2 & \alpha_1 & 0 \end{matrix}\right),
\end{equation}
can be identified with elements in the Lie algebra $so(3)$. 
Let $V_e$ be a neighbourhood of the origin in $\mathbb{R}^3$ with $|\alpha|\leq C$, a neighbourhood of $e$ in $SO(3)$ is then given by
\begin{equation}
U_e: r(\alpha)=\exp\left(\textbf{M}(\alpha)\right)\qquad \alpha\in V_e. 
\end{equation}
Let $g^*$ be another element of $SO(3)$ and let $R^*=r(\alpha)$ be its representation as a rotation matrix.
The local patch around $g^*$ is then given by
\begin{equation}
U_{g*}: R(\beta)=R^*r(\beta)\qquad \beta\in V_{e}. 
\label{eq:R}
\end{equation}
As the exponential is smooth
(\ref{eq:R}) provides a one-to-one mapping between vectors in the open ball $\sqrt{\vec{\beta}^T \vec{\beta}}<C$
and sets $U_{g*}$ if the maximal radii $C$ are small enough.
In fact, the elements along a line in $\alpha$ correspond to rotations around a given axis which give distinct outcomes
as long as the angle of rotation is less than half a turn; the constant $C$ therefore has to be less than $\pi$. Note that $r^{-1}(\alpha)=r(-\alpha)$.

Using the Levi-Civita tensor and the Einstein convention we can write
\begin{equation}
r(\alpha)_{ij} =e^{-\epsilon_{ijs}\alpha_s}.
\label{eq:r-v0}
\end{equation}
which is convenient for calculation. However, it has to be remembered that when doing so we 
connect the basis of the Lie algebra with the body frame (they have to rotate together to keep $\epsilon$ invariant).
In the main text and below we use a basis such that the angular diffusion matrix $b_{22}=k_BT\eta^{-1}$ 
entering the elliptic operator $\mathcal{L}_0$ is diagonal at the group identity element $e$.
This together with (\ref{eq:r-v0}) means that we have chosen a frame of reference for the body system,
the one for which the angular friction matrix $\eta$ is diagonal, 
$\eta=\hbox{diag}(\eta_{11},\eta_{22},\eta_{33})$.

Let now $g$ be another group element close to $g^*$. By above it can be represented by a rotation matrix which can be written either
as $r(\alpha)r(\Delta\beta)$ with some small $\Delta\beta$, or as $r(\alpha+\Delta\alpha)$, with some small $\Delta\alpha$.
This means
\begin{equation}
r(\Delta\beta) = r(-\alpha)r(\alpha+\Delta\alpha), 
\label{eq:R}
\end{equation}
which when the increments are infinitesimal gives the inverse Jacobian (upper and lower indices do not need to be distinguished since 
both $\beta$ and $\alpha$ are elements of $\mathbb{R}^3$)
\begin{equation}
\begin{split}
&(J^{-1})_{ab}=\frac{\partial\beta_{a}}{\partial\alpha_{b}}\\
&=\left(\delta_{ab}+\frac{1}{2}\epsilon_{abt}\alpha_t+\frac{1}{6}(\alpha_a\alpha_b-\alpha^2\delta_{ab})\right) + \mathcal{O}(\alpha^3).
\end{split}
\end{equation} 
The Jacobian is similarly
\begin{equation}
\begin{split}
&(J)_{ab}=\frac{\partial\alpha_{a}}{\partial\beta_{b}} \\
&=\left(\delta_{ab}-\frac{1}{2}\epsilon_{abt}\alpha_t+\frac{1}{12}(\alpha_a\alpha_b-\alpha^2\delta_{ab})\right) + \mathcal{O}(\alpha^3).
\end{split}
\end{equation} 
The volume element close to unit element is given by
\begin{equation}
d\hbox{Vol} =d\beta_1 d\beta_2 d\beta_3.
\end{equation} 
By above, the volume element in the local coordinate is therefore up to terms quadratic in $\alpha$ given by
\begin{equation}
\sqrt{g}d\alpha_1 d\alpha_2 d\alpha_3 =\left(1 - \frac{1}{12}\alpha^2\right) d\alpha_1 d\alpha_2 d\alpha_3. 
\end{equation} 

In the main text we are interested in the partial differential operator
$\mathcal{L}^{\dagger}_0$. Suppose that at the group element $g$ and in system of local coordinates $\beta$ around $g$ it
is given by a torque (drift field) $T=0$ and a constant diffusion matrix $\overline{b}$. In coordinates $\alpha'$
centered on $g^*$  $\mathcal{L}^{\dagger}_0$ is then given by
$\frac{1}{\sqrt{g'}}\partial_{{\alpha'}^a}\left(-\sqrt{g'}{T'}^a+\frac{1}{2}\partial_{{\alpha'}^b}(\sqrt{g'}{b'}^{ab})\right)$
where $T'$ and $b'$ follow from the transformations (\ref{eq:transformations}).
In the case at hand this means 
\begin{eqnarray}
{T'}^a &=& \frac{1}{2} \frac{\partial^2 {\alpha'}^a}{\partial_{\beta^p} \partial_{\beta^q}}\overline{b}^{pq}; \\
{b'}^{ab} &=&  \frac{\partial {\alpha'}^a}{\partial_{\beta^p}}  \frac{\partial {\alpha'}^b}{\partial_{\beta^q}}\overline{b}^{pq}; \\
\sqrt{g'} &=&  \det\left(\frac{\partial {\alpha'}^a}{\partial_{\beta^p}}\right)^{-1}  .
\end{eqnarray}
We recognize that the combination ${T'}^a-\frac{1}{2\sqrt{g'}}\partial_{{\alpha'}^b}(\sqrt{g'}{b'}^{ab})$ is the
H\"anggi-Klimontovich drift which transforms as a vector, and which therefore here is zero.
The operator (acting on a function $h$) is hence
\begin{equation}
\mathcal{L}^{\dagger}_0[h] = \frac{1}{2}{b'}^{ab}\frac{\partial^2 h}{\partial_{{\alpha'}^a}\partial_{{\alpha'}^b}}
+\frac{1}{2\sqrt{g'}}\partial_{{\alpha'}^a}\left(\sqrt{g'}{b'}^{ab}\right)\frac{\partial h}{\partial_{{\alpha'}^b}}.
\end{equation}
The factor 
$\partial_{\alpha^a}\left(\sqrt{g}b^{ab}\right)$ 
is of order $\mathcal{O}(\alpha)$ which means that for small enough patch it can be ignored.

\section{Invariant Method}
\label{InvariantMethodAppendix}
In this section we show the method how to averages over the orientations can be used to compute the effective diffusion tensor.
The averages over orientations are of the type
\begin{equation}
M_{ijkl}= \left<Q_{ij}\tilde{\gamma}^{-1}_{kl} \right> ,
\end{equation}
for some indices $(i,j,k,l)$ and $Q_{ij}=R^{*}_{ik}Q_{B_{kk}}(R^{*})^{-1}_{kj}$, and where the average is taken with respect
to the uniform measure (Haar measure). This can be written
\begin{eqnarray}
M_{ijkl}     &=&\frac{\partial^4}{\partial_{x_i}\partial_{y_j}\partial_{z_k}\partial_{w_l}} I(x,y,z,w), \\
I(x,y,z,w)  &=& \left<Q_{ij}\tilde{\gamma}^{-1}_{k'l'} x_{i'} y_{j'} z_{k'} w_{l'}\right>, 
\end{eqnarray}
where $x$, $y$ , $z$ and $w$ are three-dimensional vectors and $(i',j',k',l')$ are summed-over indices (Einstein convention). 
When the measure is uniform $I(x,y,z,w)$ must be an invariant of $SO(3)$ since any overall rotation of the vectors can be
brought over on the matrices $\tilde{\gamma}$ and $Q$, and all rotations of the matrices are included in the average.
By general invariant theory and by symmetry under $x\leftrightarrow y$ and  $z\leftrightarrow w$
\begin{equation}
I(x,y,z,w)  = A(x,y)(z,w)+B(x,z)(y,w)+B(x,w)(y,z) \nonumber
\end{equation}
which leads to
\begin{equation}
M_{ijkl}  = A\mathbf{1}_{ij}\mathbf{1}_{kl} +  B\mathbf{1}_{ik}\mathbf{1}_{jl} + B\mathbf{1}_{il}\mathbf{1}_{jk}, 
\end{equation}
where $A$ and $B$ are constants to be determined. We observe that
\begin{equation}
\begin{split}
& \left<Q_{ii}\tilde{\gamma}^{-1}_{jj}\right>=0 ,\\
& \left<Q_{ij}\tilde{\gamma}^{-1}_{ji}\right>=\hbox{Tr}(Q \tilde{\gamma}^{-1}), \\
\end{split}
\end{equation}
which gives the equations
\begin{equation}
\begin{split}
& 9A + 6B=0 ,\\
& 3A + 12B=\hbox{Tr}(Q \tilde{\gamma}^{-1}), \\
\end{split}
\end{equation}
The solution is
\begin{equation}
\begin{split}
&A=- \frac{\hbox{Tr}(Q\tilde{\gamma}^{-1})}{15}, \\
&B=\frac{\hbox{Tr}(Q\tilde{\gamma}^{-1})}{10}.\\
\end{split}
\end{equation}
In summary, by the invariant method we only need to compute one trace.
A similar calculation was sketched in the main body of the paper for $SU(2)$.

\bibliography{Rotation-Paper2}{}

\end{document}